\documentclass[aps,prd,nofootinbib,amsmath,amssymb,superscriptaddress,preprintnumbers,twocolumn,notitlepage]{revtex4-1}
\usepackage{lineno}
\usepackage{graphics,epsfig,subfigure}
\usepackage{color}
\usepackage{url}
\usepackage{orcidlink}
\usepackage{float}
\usepackage{dcolumn}
\usepackage{bm}
\usepackage{latexsym}
\usepackage{booktabs}
\usepackage{amsmath}
\usepackage{amssymb}
\allowdisplaybreaks[4]
\usepackage{multirow}
\newcommand{\be}{\begin{equation}}
\newcommand{\ee}{\end{equation}}
\newcommand{\bq}{\begin{eqnarray}}
\newcommand{\eq}{\end{eqnarray}}
\usepackage{graphicx}
\usepackage{amssymb}
\usepackage{epstopdf}
\usepackage{color}
\usepackage{float}
\usepackage{amsmath}
\usepackage{orcidlink}
\usepackage{latexsym}
\usepackage{amsmath}
\usepackage{amssymb}
\usepackage{amsfonts}
\usepackage{blindtext}
\usepackage{subfigure}
\usepackage{xcolor}
\usepackage[normalem]{ulem}
\begin{document}


\title{Redshift Spectroscopy as a Probe of Regular Black Holes, Black Bounces, and Scalar-Hair Compact Objects}

\author{Ali \"Ovg\"un \orcidlink{0000-0002-9889-342X}}
\email{ali.ovgun@emu.edu.tr}
\affiliation{Physics Department, Faculty of Arts and Sciences, Eastern Mediterranean University, Famagusta, 99628 North Cyprus via Mersin 10, Turkiye.}

\author{Reggie C. Pantig \orcidlink{0000-0002-3101-8591}}
\email{rcpantig@mapua.edu.ph}
\affiliation{Physics Department, School of Foundational Studies and Education, Map\'ua University, 658 Muralla St., Intramuros, Manila 1002, Philippines.}

\author{Joel Saavedra \orcidlink{0000-0002-1430-3008}}
\email{joel.saavedra@pucv.cl}
\affiliation{Instituto de F\'{\i}sica, Pontificia Universidad Cat\'olica de Valpara\'{\i}%
so, Casilla 4950, Valpara\'{\i}so, Chile.}

\begin{abstract} Motivated by [Phys. Rev. D. 107, 064019 (2023)], we develop a unified and model-independent framework for the spectroscopy of photon frequency shifts in generic static, spherically symmetric spacetimes. Working with a line element characterized by three arbitrary radial metric functions, we derive exact expressions for the conserved quantities of massive and massless probes, the conditions for circular timelike geodesics, the local-emission-angle-dependent photon impact parameter, and the corresponding local redshift and blueshift branches measured by distant static observers. The formalism is further extended to include the line-of-sight peculiar motion of the source and the local propagation of photons in a nonmagnetized cold plasma, thereby identifying the gravitational, orbital, and dispersive factors entering the frequency-shift signal under the stated assumptions. We also show that, in vacuum, the same geometric structures governing orbital spectroscopy determine the photon sphere and the shadow impact parameter whenever an external null critical orbit is present. To make the framework suitable for deformed compact-object models, we construct a perturbative expansion around Schwarzschild geometry up to second order in a dimensionless deformation parameter, obtaining explicit corrections to the orbital energy, angular momentum, emitter four-velocity, photon impact parameter, and the vacuum and plasma frequency shifts. As concrete illustrations, we apply the formalism to regular black holes from nonlinear electrodynamics, the Simpson--Visser black-bounce spacetime, and the Fisher--Janis--Newman--Winicour--Wyman geometry. \textcolor{black}{These examples show that the same local spectroscopic language can be used across regular-black-hole, black-bounce, wormhole, and scalar-supported horizonless sectors. In the Fan--Wang family, the local maximal-redshift branch is found to remain globally monotonic within the representative physical black-hole branches considered here. }\end{abstract}

\maketitle

\section{Introduction}

Black holes and other compact objects are now probed through several complementary observational channels. Gravitational-wave detections have opened a direct window onto the strong-field and dynamical regime of gravity \cite{Abbott:2016blz}, while horizon-scale very long baseline interferometry has made photon capture and light bending observationally accessible through the images of M87* and Sgr A* \cite{Akiyama:2019cqa,Akiyama:2019eap,EHT:2022wkp,EHT:2022xnr,EHT:2022yil}. Recent observations of black hole shadows, particularly those captured by the Event Horizon Telescope, have become a powerful tool for testing fundamental physics and constraining modified theories of gravity, including models involving extra dimensions, non-linear electrodynamics, scalar hair, and quantum effects \cite{Vagnozzi:2022moj, Vagnozzi:2019apd, Allahyari:2019jqz, Khodadi:2020jij, Battista:2026nsx, Wang:2025fmz}. At the same time, precision observations of stellar motion near the Galactic center have reached the sensitivity needed to detect relativistic frequency shifts, most notably in the orbit of the S2 star \cite{GRAVITY:2018ofz}. These developments make it increasingly natural to regard redshift spectroscopy, orbital dynamics, photon-sphere physics, and black-hole imaging not as separate topics, but as different observational facets of a common strong-gravity program.

Among the cleanest relativistic observables is the frequency shift of photons emitted by matter orbiting a compact object. The idea that redshift and blueshift measurements can encode information about the underlying geometry has a long history, but it was placed in a particularly elegant form by Herrera-Aguilar and Nucamendi, who expressed Kerr black-hole parameters in terms of the redshift/blueshift of photons emitted by geodesic particles \cite{HerreraAguilar:2015hua}. This framework was subsequently extended to other compact objects and rotating geometries, including boson stars, regular black holes, modified-gravity solutions, and backgrounds with frame dragging \cite{Becerril:2016ttl,Sheoran:2018wte,Becerril:2021kzb,Banerjee:2022him,Momennia:2023rpx}. Related studies have also emphasized the role of extremal redshift/blueshift configurations and line-emission spectroscopy in strong gravitational fields \cite{Gates:2020blue}.

The static, spherically symmetric sector is especially attractive as a general laboratory for this program. On the one hand, many regular black holes, effective quantum-corrected geometries, scalar-hair spacetimes, black-bounce backgrounds, and wormhole-like compact objects can be cast in this form. On the other hand, the sector remains sufficiently simple to permit exact analytic expressions, while still being broad enough to capture departures from Schwarzschild geometry in both black-hole and horizonless settings. It is also the natural setting for photon-sphere and shadow analyses, beginning with Synge's early treatment of photon escape from strongly gravitating stars and extending to modern black-hole imaging studies \cite{Synge:1966okc,Perlick:2021aok}. From this perspective, it is important to formulate the spectroscopy problem in a way that does not presuppose a particular matter source or a particular causal structure.

{\color{black}An important additional ingredient is the medium through which photons propagate. Realistic compact objects are not surrounded by perfect vacuum: ionized matter, accretion flows, and plasma environments may introduce dispersive corrections that modify optical trajectories and render lensing and shadow observables frequency dependent. Photon propagation and gravitational lensing in plasma have been studied extensively, including nonuniform plasma lensing, relativistic images in homogeneous plasma, black-hole shadows in spherically symmetric plasma environments, strong-lensing systems, Kerr plasma shadows, strong-deflection-limit lensing in inhomogeneous plasma, and the plasma imprint on photon orbits \cite{Bisnovatyi-Kogan:2010flt,Tsupko:2014lta,Perlick:2015vta,Tsupko:2012plasma,Bisnovatyi-Kogan:2017kii,Perlick:2017fio,Feleppa:2024vdk,Perlick:2023znh}. Plasma effects have also been discussed in connection with hot-spot imaging and frequency shifts in rotating backgrounds \cite{Yan:2019plasma}. The goal of the present work is complementary to full ray-tracing and shadow studies in plasma: we derive local spectroscopic frequency-shift formulas for circular emitters and show explicitly how the refractive index enters the nonradial branch of the signal.}

A particularly relevant model-specific study in this direction is the work of Fu and Zhang \cite{Fu:2022afk}, who investigated the redshift, blueshift, and gravitational redshift of photons emitted by geodesic particles orbiting a polymerized black hole, including the effects of a surrounding plasma medium. The closest general antecedent to the present work is the study of observational redshift in general spherically symmetric black holes by Mart\'inez-Valera, Momennia, and Herrera-Aguilar \cite{MartinezValera:2024obs,MartinezValera:2025add}. These developments motivate a broader framework in which orbital spectroscopy, peculiar motion, plasma corrections, photon spheres, and shadow observables can be treated within a single analytic structure and then specialized to distinct compact-object geometries, including both black-hole and horizonless sectors.

The aim of this paper is therefore to formulate a unified spectroscopic framework for photon frequency shifts in static, spherically symmetric spacetimes and to organize its specialization to representative compact-object models. Our motivation is not limited to any single geometry. Rather, we seek a common language in which regular black holes from nonlinear electrodynamics, black-bounce and wormhole geometries, and scalar-supported horizonless spacetimes can all be analyzed in a comparable way. This is also the natural setting in which to investigate how closely frequency-shift observables track photon-sphere and shadow properties across different strong-field regimes.

The paper is organized as follows. In Sec.~\ref{sec:formalism} we present the general static, spherically symmetric setup and derive the exact timelike and null geodesic relations relevant for orbital spectroscopy, including the vacuum and plasma frequency shifts. In Sec.~\ref{sec:perturbative} we develop the perturbative expansion around Schwarzschild geometry up to second order in the deformation parameter \(\delta\). In Sec.~\ref{sec:NEDexample} and Sec.~\ref{sec:spectroscopic_mapping} we study regular black holes from nonlinear electrodynamics and the associated spectroscopic map. In Sec.~\ref{sec:SVexample} we consider the Simpson-Visser black-bounce spacetime. In Sec.~\ref{sec:FJNWexample} we analyze the Fisher-Janis-Newman-Winicour-Wyman geometry. We conclude in Sec.~\ref{con}. Throughout the paper we use the metric signature \((-,+,+,+)\) and units \(G=c=\hbar=1\).

\section{Unified spectroscopic formalism in static spherically symmetric spacetimes}
\label{sec:formalism}

We begin with the most general static, spherically symmetric line element in four dimensions,
\begin{equation}
ds^2=-\alpha(r,\delta)\,dt^2+\gamma(r,\delta)\,dr^2+\beta(r,\delta)\left(d\theta^2+\sin^2\theta\,d\varphi^2\right),
\end{equation}
where \(\alpha\), \(\beta\), and \(\gamma\) are arbitrary functions of the radial coordinate, and \(\delta\) denotes a dimensionless deformation parameter controlling deviations from a chosen reference background. This parametrization is broad enough to describe a wide class of regular, effective, and beyond-GR compact-object geometries. In the undeformed limit, \(\delta=0\), one recovers Schwarzschild,
\begin{equation}
\alpha_0(r)=1-\frac{2M}{r},\qquad
\beta_0(r)=r^2,\qquad
\gamma_0(r)=\left(1-\frac{2M}{r}\right)^{-1}.
\end{equation}
Unlike Ref.~\cite{Fu:2022afk}, which focuses on a specific polymerized black-hole background, here we adopt a model-independent perturbative strategy and expand the metric functions as
\begin{eqnarray}
\alpha=\alpha_0+\delta \alpha_1+\delta^2\alpha_2+\mathcal O(\delta^3),&\qquad
\beta=\beta_0+\delta \beta_1+\delta^2\beta_2+\mathcal O(\delta^3),\qquad
\\  \gamma=\gamma_0+\delta \gamma_1+\delta^2\gamma_2+\mathcal O(\delta^3),
\end{eqnarray}
which will later be used to organize deviations from Schwarzschild in a systematic way. Because of spherical symmetry, the motion may be confined without loss of generality to the equatorial plane, \(\theta=\pi/2\). For a massive particle, stationarity and axisymmetry imply the conserved quantities
\begin{equation}
E=\alpha\,\dot t,\qquad L=\beta\,\dot\varphi,
\label{eq:conserved_massive_merged}
\end{equation}
where an overdot denotes differentiation with respect to the proper time \(\tau\). Combining these constants of motion with the normalization condition \(u^\mu u_\mu=-1\), one finds
\begin{equation}
\dot r^{\,2}=\frac{1}{\gamma}\left(\frac{E^2}{\alpha}-\frac{L^2}{\beta}-1\right).
\label{eq:radial_massive_merged}
\end{equation}
Equation \eqref{eq:radial_massive_merged} has the usual interpretation of an effective radial equation. Circular timelike geodesics are therefore singled out not only by \(\dot r=0\), but also by the vanishing of the radial derivative of the corresponding effective potential.

Imposing the circularity conditions yields
\begin{equation}
E^2=\frac{\alpha^2\beta'}{\alpha\beta'-\beta\alpha'},\qquad
L^2=\frac{\beta^2\alpha'}{\alpha\beta'-\beta\alpha'},
\label{eq:EL_merged}
\end{equation}
together with the nonvanishing components of the emitter four-velocity,
\begin{equation}
u^t=\sqrt{\frac{\beta'}{\alpha\beta'-\beta\alpha'}},\qquad
u^\varphi=\sqrt{\frac{\alpha'}{\alpha\beta'-\beta\alpha'}}.
\label{eq:utuphi_merged}
\end{equation}
These formulas already reveal an important structural point: the combination \(\alpha\beta'-\beta\alpha'\) governs both the existence of circular timelike emitters and the spectroscopic quantities constructed from them. In this sense, the kinematics of the orbit and the resulting optical signal are controlled by the same geometric kernel.

For photons, the conserved quantities are \cite{Banerjee:2022him}
\begin{equation}
E_\gamma=\alpha\,k^t,\qquad L_\gamma=\beta\,k^\varphi,
\label{eq:conserved_photon_merged}
\end{equation}
and the null condition implies
\begin{equation}
(k^r)^2=\frac{\beta E_\gamma^2-\alpha L_\gamma^2}{\alpha\beta\gamma}.
\label{eq:radial_null_merged}
\end{equation}
This relation governs the radial propagation of null geodesics and makes explicit the competition between the energy term and the angular-momentum barrier. To connect the null motion to the local emission geometry, we introduce the
impact parameter
\begin{equation}
b\equiv \frac{L_\gamma}{E_\gamma}.
\label{eq:b_def_merged}
\end{equation}
Here and in what follows we reserve \(\varphi\) for the azimuthal coordinate and
denote the local emission angle by \(\psi\).  The angle \(\psi\) is measured in
the static orthonormal frame
\begin{equation}
e_{(t)}=\alpha^{-1/2}\partial_t,\qquad
e_{(r)}=\gamma^{-1/2}\partial_r,\qquad
e_{(\varphi)}=\beta^{-1/2}\partial_\varphi .
\end{equation}
Thus,
\begin{equation}
\sin\psi=\frac{k^{(\varphi)}}{k^{(t)}} ,
\end{equation}
where \(k^{(a)}\) are tetrad components of the photon momentum.  Since
\(k^{(t)}=E_\gamma/\sqrt{\alpha}\) and
\(k^{(\varphi)}=L_\gamma/\sqrt{\beta}\), one obtains
\begin{equation}
b^2=\frac{\beta(r_{\rm e})}{\alpha(r_{\rm e})}\sin^2\psi ,
\label{eq:b2_merged}
\end{equation}
or, with the sign convention used below,
\begin{equation}
b=-\sqrt{\frac{\beta(r_{\rm e})}{\alpha(r_{\rm e})}}\sin\psi .
\label{eq:impact_merged}
\end{equation}
The special values \(\psi=0\) and \(\psi=\pm\pi/2\) correspond respectively to
radial and tangential emission in the local static frame.  These are the
directions relevant for the gravitational redshift and for the extremal
redshift/blueshift signals. The extrema of the impact parameter occur for locally tangential emission,
namely \(\psi=\pm\pi/2\). These directions play a special role because they
produce the extremal redshift and blueshift signals.

{\color{black}In vacuum, the same null sector also determines the photon sphere and the shadow. For a static, spherically symmetric spacetime, the vacuum photon sphere is obtained from the extremization of the impact parameter, which can be written as \cite{Claudel:2000yi}}
\begin{equation}
\frac{d}{dr}\left(\frac{\beta}{\alpha}\right)=0,
\label{eq:phsphere_merged}
\end{equation}
while the associated shadow impact parameter is
\begin{equation}
b_{\rm sh}^2=\left.\frac{\beta(r,\delta)}{\alpha(r,\delta)}\right|_{r=r_{\rm ph}}.
\label{eq:shadow_merged}
\end{equation}
{\color{black}Thus, in vacuum, the same background functions that determine the local orbital spectroscopy also govern the critical null structure and the shadow size. This common geometric origin is one of the main reasons why spectroscopy and imaging can be fruitfully compared within a single framework. In a dispersive plasma, however, the critical-ray and shadow structure is modified by the refractive index and by the plasma profile, as discussed below.}

We now turn to the observable frequency shift. For a photon emitted by a particle on a circular timelike orbit and detected by
a distant observer, the measured frequency shift is defined by
\cite{HerreraAguilar:2015hua}
\begin{equation}
1+z=\frac{\omega_{\rm e}}{\omega_{\rm d}}
=\frac{-k_\mu U^\mu|_{\rm e}}{-k_\mu U^\mu|_{\rm d}} .
\end{equation}
For an equatorial circular emitter,
\[
U^\mu_{\rm e}=(U^t_{\rm e},0,0,U^\varphi_{\rm e}),
\]
and for a static detector at spatial infinity,
\[
U^\mu_{\rm d}=(1,0,0,0).
\]
Using the conserved impact parameter \(b=L_\gamma/E_\gamma\), the shift becomes
\begin{equation}
1+z=U^t_{\rm e}-b\,U^\varphi_{\rm e}.
\label{eq:z_def_merged}
\end{equation}
Substituting Eqs.~\eqref{eq:utuphi_merged} and \eqref{eq:impact_merged}, one obtains
\begin{widetext}
\begin{equation}
1+z=
\sqrt{\frac{\beta'(r_{\rm e})}
{\alpha(r_{\rm e})\beta'(r_{\rm e})-\beta(r_{\rm e})\alpha'(r_{\rm e})}}
+
\sin\psi\,
\sqrt{
\frac{\beta(r_{\rm e})\alpha'(r_{\rm e})}
{\alpha(r_{\rm e})\left[\alpha(r_{\rm e})\beta'(r_{\rm e})
-\beta(r_{\rm e})\alpha'(r_{\rm e})\right]}
}.
\label{eq:z_explicit_merged}
\end{equation}
\end{widetext}
{\color{black}The special values \(\psi=0\) and \(\psi=\pm\pi/2\) correspond respectively to radial and tangential emission in the local static frame. The radial direction removes the longitudinal Doppler contribution proportional to the photon impact parameter, whereas the locally tangential directions define the two limiting local spectroscopic branches. We shall refer to these as the local redshift/blueshift branches. They should not be confused with extrema of an observed line profile unless the emitter--observer ray-connection problem is also solved.}

{\color{black}For radial emission, \(b=0\), and the longitudinal Doppler term proportional to
the photon impact parameter vanishes. The resulting branch is therefore
\begin{equation}
1+z_{\rm rad}
=
\frac{U^t_{\rm e}}{U^t_{\rm d}}
=
U^t_{\rm e},
\label{eq:zrad_merged}
\end{equation}
for a static detector at infinity. We stress that this is not the redshift of a
static emitter. Since the source is a circularly orbiting massive particle,
\(1+z_{\rm rad}\) still contains the transverse-Doppler or orbital
time-dilation contribution. More explicitly, one may write
\begin{equation}
1+z_{\rm rad}
=
\frac{\Gamma_{\rm orb}}{\sqrt{\alpha(r_{\rm e})}},
\qquad
\Gamma_{\rm orb}
=
\sqrt{\alpha(r_{\rm e})}\,U^t_{\rm e}.
\label{eq:grav_transverse_split}
\end{equation}
The factor \(1/\sqrt{\alpha(r_{\rm e})}\) is the gravitational redshift of a
static emitter at \(r_{\rm e}\), while \(\Gamma_{\rm orb}\) encodes the
transverse-Doppler/time-dilation effect due to the circular motion of the
emitter. Thus radial emission isolates the branch with no longitudinal Doppler
contribution, but it does not remove the orbital time-dilation factor.}

If the compact-object system has a line-of-sight peculiar velocity relative to
the distant observer, the total shift acquires an additional special-relativistic
Doppler factor.  We denote the dimensionless line-of-sight velocity by
\(\upsilon_0=v_{\rm los}/c\), with \(\upsilon_0>0\) for recession.  Then
\begin{equation}
\Xi\equiv \sqrt{\frac{1+\upsilon_0}{1-\upsilon_0}} .
\label{eq:Xi_merged}
\end{equation}
Equivalently, if one uses the observed peculiar redshift \(z_{\rm pec}\), then
\(\Xi=1+z_{\rm pec}\).  In the following we use \(\upsilon_0\), rather than
calling it a redshift, to avoid confusion. The full observed shift can then be written as
\begin{equation}
1+z_{\rm tot}=\Xi(1+z),
\label{eq:ztot_compact_merged}
\end{equation}
or explicitly,
\begin{widetext}
\begin{equation}
1+z_{\rm tot}
=
\Xi\left[
\sqrt{\frac{\beta'(r_{\rm e})}
{\alpha(r_{\rm e})\beta'(r_{\rm e})-\beta(r_{\rm e})\alpha'(r_{\rm e})}}
+
\sin\psi\,
\sqrt{
\frac{\beta(r_{\rm e})\alpha'(r_{\rm e})}
{\alpha(r_{\rm e})\left[\alpha(r_{\rm e})\beta'(r_{\rm e})
-\beta(r_{\rm e})\alpha'(r_{\rm e})\right]}
}
\right].
\label{eq:ztot_explicit_merged}
\end{equation}
\end{widetext}

We stress that \(\psi\) is a local emission angle at the source.  For a fixed
detector position, not every value of \(\psi\) corresponds to a photon received
by the observer.  The allowed values are selected by the emitter-observer
boundary-value problem for null rays, and in the strong-field regime more than
one ray may connect the same emitter and detector.  Solving this problem requires
ray tracing and is beyond the scope of the present analytic treatment.  The
formulas derived here should therefore be interpreted as local spectroscopic
building blocks and as analytic expressions for the extremal redshift/blueshift
directions, not as a complete observed line-profile calculation.

This expression is the central local spectroscopic observable of the paper:
within the stated assumptions of a circular equatorial emitter and a static
asymptotic detector, it is exact, model independent, and immediately ready to be
specialized to concrete compact-object geometries.

{\color{black}Three local emission directions are of particular interest, namely \(\psi=0\) and \(\psi=\pm\pi/2\). The first corresponds to the radial emission branch, while the latter two correspond to the two locally tangential spectroscopic branches. These are useful local analytic diagnostics, but they are not automatically the extrema of an observed spectral line unless the emitter--observer null boundary-value problem is also specified. It is therefore natural to introduce the compact local spectroscopic triplet}
\begin{equation}
R\equiv 1+z_{{\rm tot},1},\qquad
S\equiv 1+z_{{\rm tot},2},\qquad
T\equiv \left(1+z_{{\rm tot},3}\right)^2.
\label{eq:RST_merged}
\end{equation}
These quantities provide a concise and model-independent way to encode the spectroscopic information carried by the spacetime.

The same logic extends to a nonmagnetized cold plasma, but in this case one must
distinguish carefully between the wave covector \(p_\mu\) and the ray tangent
\(K^\mu=dx^\mu/d\lambda\).  Photon propagation in a cold plasma is governed by
the Hamiltonian \cite{Rogers:2015dla,Perlick:2015vta,Atamurotov:2015nra}
\begin{equation}
H=\frac12\left[g^{\mu\nu}p_\mu p_\nu+\omega_{\rm p}^2(r)\right]=0 .
\label{eq:H_merged}
\end{equation}
The conserved quantities associated with stationarity and spherical symmetry are
\begin{equation}
E_\gamma=-p_t,\qquad L_\gamma=p_\varphi .
\label{eq:conserved_plasma_merged}
\end{equation}
Hamilton's equations give
\begin{equation}
K^t=\frac{\partial H}{\partial p_t}
=g^{tt}p_t=\frac{E_\gamma}{\alpha},
\qquad
K^\varphi=\frac{\partial H}{\partial p_\varphi}
=g^{\varphi\varphi}p_\varphi=\frac{L_\gamma}{\beta}.
\label{eq:hamilton_plasma_components}
\end{equation}
This point is important: the factor \(n^2\) does not appear in the conserved
energy relation.  The refractive index enters instead through the local
dispersion relation.

For a static plasma with four-velocity
\begin{equation}
V^\mu=\alpha^{-1/2}\delta^\mu_t ,
\end{equation}
the photon frequency measured in the plasma rest frame is
\begin{equation}
\omega=-p_\mu V^\mu=\frac{E_\gamma}{\sqrt{\alpha}} .
\end{equation}
The refractive index is therefore
\begin{equation}
n^2(r)=1-\frac{\omega_{\rm p}^2(r)}{\omega^2}
=1-\frac{\alpha(r)\omega_{\rm p}^2(r)}{E_\gamma^2}.
\label{eq:n2_merged}
\end{equation}
For a power-law density profile,
\begin{equation}
\omega_{\rm p}^2(r)=\frac{4\pi e^2}{m}N(r),
\qquad
N(r)=\frac{N_0}{r^h},
\label{eq:plasma_profile_merged}
\end{equation}
one obtains, at fixed frequency \(E_\gamma=\omega_0\) measured at infinity,
\begin{equation}
n^2(r)=1-\alpha(r,\delta)\frac{k}{r^h},
\qquad
k=\frac{4\pi e^2N_0}{m\omega_0^2}.
\label{eq:n2_profile_merged}
\end{equation}

The equatorial plasma dispersion relation follows from \(H=0\):
\begin{equation}
-\frac{E_\gamma^2}{\alpha}
+\gamma\left(K^r\right)^2
+\frac{L_\gamma^2}{\beta}
+\omega_{\rm p}^2(r)=0 .
\end{equation}
Using Eq.~\eqref{eq:n2_merged}, this may be written as
\begin{equation}
-\frac{n^2(r)}{\alpha}E_\gamma^2
+\gamma\left(K^r\right)^2
+\frac{L_\gamma^2}{\beta}=0 .
\label{eq:null_plasma_merged}
\end{equation}

The local spatial wave number in the plasma rest frame is \(n\omega\).  Hence the
plasma-corrected impact parameter, written in terms of the physical emission
angle \(\psi\), is
\begin{equation}
\hat b\equiv \frac{L_\gamma}{E_\gamma}
=
-n(r_{\rm e})
\sqrt{\frac{\beta(r_{\rm e})}{\alpha(r_{\rm e})}}
\sin\psi .
\label{eq:bhat_merged}
\end{equation}
The frequency shift is still computed from the wave covector:
\begin{equation}
1+\hat z=
\frac{-p_\mu U^\mu|_{\rm e}}{-p_\mu U^\mu|_{\rm d}}.
\end{equation}
For a circular equatorial emitter and a static detector at infinity this gives
\begin{equation}
1+\hat z=U^t_{\rm e}-\hat b\,U^\varphi_{\rm e}.
\end{equation}
Therefore the total plasma-corrected shift is
\begin{widetext}
\begin{equation}
1+\hat z_{\rm tot}
=
\Xi\left[
\sqrt{\frac{\beta'(r_{\rm e})}
{\alpha(r_{\rm e})\beta'(r_{\rm e})-\beta(r_{\rm e})\alpha'(r_{\rm e})}}
+
n(r_{\rm e})\sin\psi\,
\sqrt{
\frac{\beta(r_{\rm e})\alpha'(r_{\rm e})}
{\alpha(r_{\rm e})\left[\alpha(r_{\rm e})\beta'(r_{\rm e})
-\beta(r_{\rm e})\alpha'(r_{\rm e})\right]}
}
\right].
\label{eq:ztot_plasma_merged}
\end{equation}
\end{widetext}

{\color{black}The above plasma extension should be understood as a local spectroscopic extension. A full calculation of plasma-modified shadows or observed line profiles requires solving the dispersive ray problem between the emitter and the observer. In a cold plasma the critical-ray structure is generally frequency dependent. At fixed photon energy \(E_\gamma=\omega_0\) measured at infinity, the radial turning-point condition gives the plasma impact parameter \begin{equation} b_{\rm pl}^2(r;\omega_0) = \frac{\beta(r)}{\alpha(r)} \left[ 1-\frac{\alpha(r)\omega_{\rm p}^2(r)}{\omega_0^2} \right]. \label{eq:bpl_critical} \end{equation} Hence a plasma-modified circular critical ray, when it exists, is determined by \begin{equation} \frac{d}{dr} \left\{ \frac{\beta(r)}{\alpha(r)} \left[ 1-\frac{\alpha(r)\omega_{\rm p}^2(r)}{\omega_0^2} \right] \right\} =0. \label{eq:plasma_critical_condition} \end{equation} Equivalently, \begin{equation} \frac{d}{dr} \left[ \frac{\beta(r)}{\alpha(r)} - \frac{\beta(r)\omega_{\rm p}^2(r)}{\omega_0^2} \right] =0. \label{eq:plasma_critical_condition_equiv} \end{equation} Equations \eqref{eq:bpl_critical}--\eqref{eq:plasma_critical_condition_equiv} reduce to the vacuum photon-sphere condition \(d(\beta/\alpha)/dr=0\) when \(\omega_{\rm p}\to0\). In the present paper we do not perform full plasma ray tracing or compute plasma-modified shadows; instead, we use the plasma formalism to identify how the refractive index modifies the local nonradial spectroscopic branch.}
The plasma affects the nonradial optical contribution through \(n(r_{\rm e})\).
{\color{black}For radial emission, \(\psi=0\), one has \(\hat b=0\), and the radial emission branch is unchanged by the plasma. This statement refers to the local frequency shift; it does not imply that plasma leaves the full emitter--observer ray connection or line profile unchanged.}

In analogy with the vacuum case, the plasma-corrected spectroscopic triplet is
\begin{equation}
\hat R\equiv 1+\hat z_{{\rm tot},1},\qquad
\hat S\equiv 1+\hat z_{{\rm tot},2},\qquad
\hat T\equiv \left(1+\hat z_{{\rm tot},3}\right)^2.
\label{eq:RST_plasma_merged}
\end{equation}

\section{Perturbative expansion around Schwarzschild}
\label{sec:perturbative}

The exact expressions derived in the previous sections are model independent and
apply to any static, spherically symmetric geometry. In many applications,
however, the spacetime of interest is naturally interpreted as a small deformation
of Schwarzschild. This motivates a perturbative treatment in which the metric
functions are expanded in powers of a dimensionless deformation parameter
\(\delta\),
\begin{align}
\alpha(r,\delta)&=\alpha_0(r)+\delta\,\alpha_1(r)+\delta^2\,\alpha_2(r)+\mathcal{O}(\delta^3),\\
\beta(r,\delta)&=\beta_0(r)+\delta\,\beta_1(r)+\delta^2\,\beta_2(r)+\mathcal{O}(\delta^3),\\
\gamma(r,\delta)&=\gamma_0(r)+\delta\,\gamma_1(r)+\delta^2\,\gamma_2(r)+\mathcal{O}(\delta^3),
\end{align}
with the Schwarzschild background
\begin{equation}
\alpha_0(r)=1-\frac{2M}{r},\qquad
\beta_0(r)=r^2,\qquad
\gamma_0(r)=\left(1-\frac{2M}{r}\right)^{-1}.
\end{equation}
This framework is sufficiently broad to accommodate regular black holes,
effective quantum corrections, scalar-hair backgrounds, and other static
spherical deformations.

The role of the perturbative expansion is not to replace the exact formulas,
which are available in the static spherical sector, but to identify which
combinations of deformation functions are probed by frequency-shift spectroscopy
at a given order.  In this sense the expansion provides a response theory around
Schwarzschild geometry: it separates the metric deformations entering the
timelike orbital sector from those entering the optical sector, and it makes
explicit the degeneracies that may occur when different metric functions produce
the same spectroscopic signal.

The central geometric quantity entering the spectroscopy is
\begin{equation}
\mathcal{D}(r,\delta)\equiv \alpha\,\beta'-\beta\,\alpha'
=\mathcal{D}_0+\delta \mathcal{D}_1+\delta^2 \mathcal{D}_2+\mathcal{O}(\delta^3),
\label{eq:D_definition_main}
\end{equation}
where the explicit coefficients \(\mathcal{D}_0,\mathcal{D}_1,\mathcal{D}_2\)
are listed in Appendix~\ref{app:perturbative_details}. This object plays the role
of the universal kernel of the perturbative construction: it controls the
circular timelike geodesics, the emitter four-velocity, and consequently both the
vacuum and plasma frequency shifts.

To propagate the expansion systematically through the observables, it is
convenient to use the generic quotient formula
\begin{equation}
\frac{N_0+\delta N_1+\delta^2 N_2}{D_0+\delta D_1+\delta^2 D_2}
=
Q_0+\delta Q_1+\delta^2 Q_2+\mathcal{O}(\delta^3),
\label{eq:quotient_generic_main}
\end{equation}
together with the corresponding square-root expansion
\begin{eqnarray}
\sqrt{Q_0+\delta Q_1+\delta^2 Q_2}
=
\sqrt{Q_0}
+\delta\,\frac{Q_1}{2\sqrt{Q_0}}
\notag \\+\delta^2\left(\frac{Q_2}{2\sqrt{Q_0}}-\frac{Q_1^2}{8Q_0^{3/2}}\right)
+\mathcal{O}(\delta^3).
\label{eq:sqrt_generic_main}
\end{eqnarray}
The explicit coefficients \(Q_0,Q_1,Q_2\) are given in
Appendix~\ref{app:perturbative_details}. These two elementary expansions are the
only algebraic ingredients needed to derive the perturbative form of all
spectroscopic quantities.

The exact circular-orbit expressions
\begin{equation}
E^2=\frac{\alpha^2\beta'}{\alpha\beta'-\beta\alpha'},
\qquad
L^2=\frac{\beta^2\alpha'}{\alpha\beta'-\beta\alpha'}
\end{equation}
admit the perturbative expansions
\begin{eqnarray}
E^2=E_{(0)}^2+\delta E_{(1)}^2+\delta^2 E_{(2)}^2+\mathcal{O}(\delta^3),
\qquad  \\ \notag
L^2=L_{(0)}^2+\delta L_{(1)}^2+\delta^2 L_{(2)}^2+\mathcal{O}(\delta^3).
\label{eq:EL_pert_main}
\end{eqnarray}
Likewise, the emitter four-velocity components
\begin{equation}
u^t=\sqrt{\frac{\beta'}{\mathcal{D}}},
\qquad
u^\varphi=\sqrt{\frac{\alpha'}{\mathcal{D}}},
\end{equation}
take the form
\begin{align}
u^t&=u^t_{(0)}+\delta u^t_{(1)}+\delta^2 u^t_{(2)}+\mathcal{O}(\delta^3),\\
u^\varphi&=u^\varphi_{(0)}+\delta u^\varphi_{(1)}+\delta^2 u^\varphi_{(2)}+\mathcal{O}(\delta^3).
\label{eq:upert_main}
\end{align}
The explicit coefficients of \(E^2\), \(L^2\), \(u^t\), and \(u^\varphi\) are
collected in Appendix~\ref{app:perturbative_details}. Their structure makes clear
which combinations of deformation functions enter the timelike orbital
spectroscopy at each order.

For null rays emitted at the local angle \(\psi\), measured in the static
orthonormal frame, the exact impact parameter is
\begin{equation}
b
=
-\sqrt{\frac{\beta(r,\delta)}{\alpha(r,\delta)}}\,\sin\psi .
\label{eq:b_exact_main}
\end{equation}
The perturbative expansion then reads
\begin{equation}
b=b_{(0)}+\delta b_{(1)}+\delta^2 b_{(2)}+\mathcal{O}(\delta^3),
\label{eq:b_pert_main}
\end{equation}
with explicit coefficients given in Appendix~\ref{app:perturbative_details}.
In this local-angle formulation, the impact parameter depends directly on the
ratio \(\beta/\alpha\). The radial metric function \(\gamma\) enters the radial
ray propagation, but not the locally measured emission-angle relation.

For an asymptotically flat observer at infinity,
\begin{equation}
1+z=u^t_{\rm e}-b\,u^\varphi_{\rm e},
\end{equation}
and the redshift therefore expands as
\begin{equation}
1+z=\mathcal{Z}_0+\delta \mathcal{Z}_1+\delta^2 \mathcal{Z}_2+\mathcal{O}(\delta^3),
\label{eq:z_expansion_main}
\end{equation}
where
\begin{align}
\mathcal{Z}_0&=u^t_{(0)}-b_{(0)}u^\varphi_{(0)},\\
\mathcal{Z}_1&=u^t_{(1)}-b_{(0)}u^\varphi_{(1)}-b_{(1)}u^\varphi_{(0)},\\
\mathcal{Z}_2&=u^t_{(2)}-b_{(0)}u^\varphi_{(2)}-b_{(1)}u^\varphi_{(1)}-b_{(2)}u^\varphi_{(0)}.
\label{eq:Z_coefficients_main}
\end{align}
{\color{black}For radial emission, \(b=0\), the longitudinal Doppler term vanishes and the radial emission branch becomes \begin{equation} 1+z_{\rm rad} = u^t_{(0)} +\delta u^t_{(1)} +\delta^2 u^t_{(2)} +\mathcal{O}(\delta^3). \end{equation} As in the exact treatment, this branch contains both the static gravitational redshift factor and the transverse-Doppler/time-dilation contribution due to the circular motion of the emitter.}
If the compact-object system has line-of-sight peculiar velocity
\(\upsilon_0=v_{\rm los}/c\), we define
\begin{equation}
\Xi\equiv \sqrt{\frac{1+\upsilon_0}{1-\upsilon_0}} .
\end{equation}
so that the total shift reads
\begin{equation}
z_{\rm tot}
=
\Xi\left(\mathcal{Z}_0+\delta \mathcal{Z}_1+\delta^2 \mathcal{Z}_2\right)-1
+\mathcal{O}(\delta^3).
\label{eq:ztot_pert_main}
\end{equation}
Equation \eqref{eq:ztot_pert_main} provides the desired order-by-order
deformation of the full spectroscopic observable.

In particular, the extremal observational triplet
\begin{equation}
R\equiv 1+z_{{\rm tot},1},\qquad
S\equiv 1+z_{{\rm tot},2},\qquad
T\equiv \left(1+z_{{\rm tot},3}\right)^2
\end{equation}
inherits a perturbative expansion of the form
\begin{align}
R&=R_{(0)}+\delta R_{(1)}+\delta^2 R_{(2)}+\mathcal{O}(\delta^3),\\
S&=S_{(0)}+\delta S_{(1)}+\delta^2 S_{(2)}+\mathcal{O}(\delta^3),\\
T&=T_{(0)}+\delta T_{(1)}+\delta^2 T_{(2)}+\mathcal{O}(\delta^3),
\end{align}
whose explicit coefficients are listed in
Appendix~\ref{app:perturbative_details}. These quantities provide a natural
starting point for model-dependent inverse problems and parameter estimation.

The same strategy applies in the presence of plasma. Since
\begin{equation}
n^2(r)=1-\alpha(r,\delta)\frac{k}{r^h},
\end{equation}
the refractive index admits the expansion
\begin{equation}
n=n_{(0)}+\delta n_{(1)}+\delta^2 n_{(2)}+\mathcal{O}(\delta^3),
\label{eq:n_expansion_main}
\end{equation}
with the explicit coefficients given in Appendix~\ref{app:perturbative_details}.
Because \(\hat b=n\,b\), the plasma-corrected impact parameter becomes
\begin{equation}
\hat b=\hat b_{(0)}+\delta \hat b_{(1)}+\delta^2 \hat b_{(2)}+\mathcal{O}(\delta^3),
\end{equation}
and the corresponding redshift takes the form
\begin{equation}
1+\hat z
=
\hat{\mathcal{Z}}_0+\delta \hat{\mathcal{Z}}_1+\delta^2 \hat{\mathcal{Z}}_2
+\mathcal{O}(\delta^3).
\label{eq:zhat_expansion_main}
\end{equation}
Hence the total plasma-corrected spectroscopic observable is
\begin{equation}
1+\hat z_{\rm tot}
=
\Xi\left(\hat{\mathcal{Z}}_0+\delta \hat{\mathcal{Z}}_1+\delta^2 \hat{\mathcal{Z}}_2\right)
+\mathcal{O}(\delta^3).
\label{eq:zhat_total_main}
\end{equation}
As in the exact analysis, the radial-emission branch remains
unchanged by the plasma at every perturbative order, because \(\hat b=0\) for
\(\psi=0\). The plasma therefore modifies only the nonradial optical sector of
the spectroscopic signal.
\section{Example 1: Regular black holes from nonlinear electrodynamics}
\label{sec:NEDexample}

As a first explicit application of the general framework, we consider a broad
family of regular black holes generated by nonlinear electrodynamics (NED). This
class is especially useful because it furnishes a controlled deformation of the
Schwarzschild geometry while remaining analytically tractable and encompassing
several well-known regular solutions as special cases. In particular, it allows
one to connect the general spectroscopy formalism directly to Bardeen-like and
Hayward-like geometries within a single unified metric ansatz.

We begin from the Einstein-NED action
\begin{equation}
S=\frac{1}{16\pi}\int d^4x\,\sqrt{-g}\,\left[R-\mathcal{L}(F)\right],
\label{eq:NED_action}
\end{equation}
where \(R\) is the Ricci scalar, \(g\) is the determinant of the metric,
\(F\equiv F_{\mu\nu}F^{\mu\nu}\), and
\begin{equation}
F_{\mu\nu}=\partial_\mu A_\nu-\partial_\nu A_\mu
\end{equation}
is the electromagnetic field strength tensor. A wide class of static,
spherically symmetric, magnetically charged regular black holes can then be
written in the form
\begin{equation}
ds^2=-f(r)\,dt^2+\frac{dr^2}{f(r)}+r^2\,d\Omega^2,
\label{eq:NED_metric}
\end{equation}
with lapse function \cite{Toshmatov:2019gxg}
\begin{equation}
f(r)=1-\frac{2M r^{\mu-1}}{\left(r^\nu+q_m^\nu\right)^{\mu/\nu}}.
\label{eq:NED_f}
\end{equation}
Here \(M\) is the mass parameter, \(q_m\) is the magnetic charge parameter, and
\(\mu,\nu>0\) characterize the nonlinear-electrodynamic deformation. This form is
particularly convenient because it is asymptotically Schwarzschild and, for
appropriate parameter values, regularizes the central singularity.

The geometry \eqref{eq:NED_metric} is immediately mapped to the general
formalism of the previous sections through the identification
\begin{equation}
\alpha(r)=f(r),\qquad
\beta(r)=r^2,\qquad
\gamma(r)=\frac{1}{f(r)}.
\label{eq:NED_identification}
\end{equation}
Accordingly, all general formulas for circular geodesics, photon impact
parameters, and frequency shifts can be specialized to this example in closed
form. The horizons are determined by the roots of$
f(r_h)=0.$ For nonextremal configurations there are generally two horizons, an inner Cauchy
horizon and an outer event horizon. The extremal configuration is obtained by
the double-root conditions $f(r_h)=0,$ and  $f'(r_h)=0.$ These conditions are useful because they identify the boundary in parameter space
at which the horizon structure changes qualitatively.

The derivative of the lapse function is
\begin{equation}
f'(r)
=
-\frac{2M r^{\mu-2}}{\left(r^\nu+q_m^\nu\right)^{\mu/\nu+1}}
\left[
(\mu-1)\left(r^\nu+q_m^\nu\right)-\mu r^\nu
\right].
\label{eq:NED_fprime}
\end{equation}
This derivative enters directly in the conditions for circular timelike orbits,
photon spheres, and the observable redshift/blueshift formulas derived below.

At large radius, the lapse function admits the expansion
\begin{equation}
f(r)
=
1-\frac{2M}{r}
+\frac{2M\mu}{\nu}\frac{q_m^\nu}{r^{\nu+1}}
-\frac{M\mu(\mu+\nu)}{\nu^2}\frac{q_m^{2\nu}}{r^{2\nu+1}}
+\mathcal{O}\!\left(r^{-3\nu-1}\right).
\label{eq:NED_large_r}
\end{equation}
This makes explicit that the spacetime is asymptotically Schwarzschild and shows
how the magnetic charge enters as a controlled deformation at finite radius.
Equation \eqref{eq:NED_large_r} also makes clear why this family provides a
natural testing ground for the perturbative framework developed in
Sec.~\ref{sec:perturbative}.

The line element \eqref{eq:NED_metric} interpolates among several standard
regular-black-hole geometries. The subclass \(\nu=1\) is often referred to as
the Maxwellian regular black hole, \(\nu=2\) yields the Bardeen-like family, and
\(\nu=3\) yields the Hayward-like family. In addition, regularity at the center
is achieved for
\begin{equation}
\mu\ge 3.
\label{eq:mu_regular}
\end{equation}
For phenomenological purposes one may either keep \(\mu\) and \(\nu\) arbitrary
or, for definiteness, focus on the minimal regular choice \(\mu=3\), which
already captures the essential physics of the regular core.

We next specialize the circular timelike geodesics. Since \(\beta=r^2\), one has
\(\beta'=2r\), and the denominator appearing in the generic formulas becomes
\begin{equation}
\alpha\beta'-\beta\alpha'
=
2r\,f(r)-r^2 f'(r)
=
r\left[2f(r)-r f'(r)\right].
\label{eq:NED_D}
\end{equation}
This quantity plays the same role here as in the model-independent formalism: it
controls the existence of circular timelike emitters and governs the
spectroscopic observables through the emitter four-velocity.

The conserved energy and angular momentum per unit rest mass are therefore
\begin{equation}
E^2=\frac{2f(r)^2}{2f(r)-r f'(r)},
\label{eq:NED_E2}
\end{equation}
and
\begin{equation}
L^2=\frac{r^3 f'(r)}{2f(r)-r f'(r)}.
\label{eq:NED_L2}
\end{equation}
These expressions show explicitly that circular timelike motion requires
\begin{equation}
2f(r)-r f'(r)>0,
\label{eq:NED_timelike_cond1}
\end{equation}
while the positivity of \(L^2\) further requires
\begin{equation}
f'(r)>0.
\label{eq:NED_timelike_cond2}
\end{equation}
{\color{black}In the plots and analytic maps below we display the full circular-timelike domain whenever it is useful for diagnosing the geometry. This domain can include unstable circular geodesics. Therefore, the divergence of a local spectroscopic branch near the circular-orbit boundary should be interpreted as a formal geometric feature of circular-orbit spectroscopy. For an astrophysical disk or long-lived emitter interpretation, one must further restrict to the stable branch satisfying the usual condition \(V_{\rm eff}''(r)>0\), with the sign convention appropriate to the chosen effective potential.}

The corresponding components of the emitter four-velocity are
\begin{equation}
u^t
=
\sqrt{\frac{2}{2f(r)-r f'(r)}},
\label{eq:NED_ut}
\end{equation}
and
\begin{equation}
u^\varphi
=
\sqrt{\frac{f'(r)}{r\left[2f(r)-r f'(r)\right]}}.
\label{eq:NED_uphi}
\end{equation}
These are the basic kinematical ingredients entering the frequency-shift
observables. In particular, \(u^t\) controls the radial-emission branch of the
signal, whereas \(u^\varphi\) governs the orbital Doppler contribution.

\subsection{Photon impact parameter, redshift, and shadow}

For photons emitted at a local angle \(\psi\), measured in the static orthonormal
frame at the emission point, the impact parameter becomes
\begin{equation}
b(r,\psi)
=
-\frac{r}{\sqrt{f(r)}}\sin\psi .
\label{eq:NED_b}
\end{equation}
This expression makes explicit how the local emission geometry and the background
lapse function combine to determine the optical trajectory seen by a distant
observer.

For an asymptotically flat observer at infinity, the observed vacuum frequency
shift is
\begin{equation}
1+z
=
u^t_{\rm e}-b\,u^\varphi_{\rm e},
\label{eq:NED_redshift}
\end{equation}
or, after substituting Eqs. \eqref{eq:NED_ut}, \eqref{eq:NED_uphi}, and
\eqref{eq:NED_b},
\begin{equation}
1+z
=
\sqrt{\frac{2}{2f(r_{\rm e})-r_{\rm e}f'(r_{\rm e})}}
+
\sin\psi\,
\sqrt{
\frac{r_{\rm e}f'(r_{\rm e})}
{f(r_{\rm e})\left[2f(r_{\rm e})-r_{\rm e}f'(r_{\rm e})\right]}
}.
\label{eq:NED_redshift_explicit}
\end{equation}
Equation \eqref{eq:NED_redshift_explicit} shows that the spectroscopic signal is
fully controlled by the lapse function and its derivative at the emitter radius.
It therefore provides a particularly transparent realization of the general
formalism in a concrete regular-black-hole background.

If the line-of-sight peculiar velocity of the compact-object system relative to
the detector is included, then
\begin{equation}
\Xi\equiv \sqrt{\frac{1+\upsilon_0}{1-\upsilon_0}},
\end{equation}
and the total shift becomes
\begin{equation}
1+z_{\rm tot}
=
\Xi\,(1+z).
\label{eq:NED_ztot}
\end{equation}
Thus, the NED example retains the same clean multiplicative structure between
local strong-field effects and the overall peculiar-motion correction.

The distinguished observational directions are exactly the same as in the
general discussion:
\begin{equation}
\psi=0,\qquad \psi=\pm \frac{\pi}{2}.
\end{equation}
{\color{black}The case \(\psi=0\) corresponds to the radial emission branch and removes the longitudinal Doppler contribution proportional to \(b\). The directions \(\psi=\pm\pi/2\) define the two locally tangential spectroscopic branches. These local branches are useful analytic diagnostics of the spacetime, but they should not be identified with extrema of an observed line profile unless the emitter--observer ray connection is also specified.}

The photon sphere is determined by the general condition
\begin{equation}
\frac{d}{dr}\left(\frac{r^2}{f(r)}\right)=0,
\end{equation}
which is equivalent to
\begin{equation}
r f'(r)-2f(r)=0.
\label{eq:NED_phsphere}
\end{equation}
Once \(r_{\rm ph}\) is determined, the shadow impact parameter is
\begin{equation}
b_{\rm sh}
=
\frac{r_{\rm ph}}{\sqrt{f(r_{\rm ph})}}.
\label{eq:NED_shadow}
\end{equation}
This makes particularly transparent the link between orbital spectroscopy and
shadow observables: both are governed by the same lapse function \(f(r)\) and
its derivatives. In this sense, the NED family provides an especially clean
benchmark for joint spectroscopy-imaging analyses.

Finally, this example also illustrates how the perturbative framework of
Sec.~\ref{sec:perturbative} can be implemented in practice. Since the large-radius
expansion \eqref{eq:NED_large_r} exhibits the charge-dependent terms as
controlled deviations from Schwarzschild, one may introduce a small
dimensionless deformation parameter of the form
\begin{equation}
\delta\sim \left(\frac{q_m}{M}\right)^\nu,
\end{equation}
and organize the metric functions as
\begin{equation}
\alpha(r,\delta)=f(r),\qquad
\beta(r,\delta)=r^2,\qquad
\gamma(r,\delta)=\frac{1}{f(r)}.
\end{equation}
In this way, the general perturbative machinery developed earlier is seen to
apply directly to a physically relevant class of regular black-hole geometries.

The metric \eqref{eq:NED_f} belongs to the Fan-Wang class of regular black-hole
solutions generated by general relativity coupled to nonlinear electrodynamics
\cite{Fan:2016hvf}.  The Fan-Wang family is therefore an instructive test bed
for the general spectroscopy formalism.  It is broad enough to include several
standard regular black holes, yet simple enough to admit closed expressions for
the circular geodesics, the redshift/blueshift observables, the photon sphere,
and the shadow.  This combination of generality and analytic control makes it
particularly useful for testing the phenomenological reach of the framework
developed in this paper.

\subsection{Spectroscopic mapping of the NED family}
\label{sec:spectroscopic_mapping}

The Fan-Wang family also provides a useful setting in which one can analyze the invertibility of the spectroscopic map. In the present context, the relevant observable is the maximal redshift, obtained for the local emission
angle \(\psi=\pi/2\). Using Eqs. \eqref{eq:NED_ut}, \eqref{eq:NED_uphi}, and
\eqref{eq:NED_ztot}, one finds
\begin{equation}
1+z_{\max}(r)
=
\Xi\left[
\sqrt{\frac{2}{\mathcal{D}(r)}}
+
\sqrt{\frac{r f'(r)}{f(r)\,\mathcal{D}(r)}}
\right],
\label{eq:zmax_def}
\end{equation}
where
\begin{equation}
\mathcal{D}(r)\equiv 2f(r)-r f'(r),
\label{eq:Dmap}
\end{equation}
and, as before,
\begin{equation}
\Xi\equiv \sqrt{\frac{1+\upsilon_0}{1-\upsilon_0}}.
\end{equation}
Since \(\Xi\) is a positive constant for fixed line-of-sight velocity
\(\upsilon_0\), the
monotonicity properties of \(z_{\max}(r)\) are entirely governed by the radial
dependence of the bracketed term in Eq. \eqref{eq:zmax_def}.

A first important observation is that the condition $\mathcal{D}(r)=0$
defines a geometric boundary of the spectroscopic description. Indeed,
Eq. \eqref{eq:zmax_def} diverges as \(\mathcal{D}(r)\to 0^+\), so this surface
does not represent a genuine extremum of the observable but rather the boundary
beyond which circular timelike spectroscopy ceases to be well defined. Hence the
physically relevant spectroscopic domain is restricted by
$\mathcal{D}(r)>0,$
supplemented, when required, by the conditions \(f(r)>0\), \(f'(r)>0\),
\(E^2>0\), and \(L^2>0\). If one wishes to model actual orbiting emitters, one
should in addition restrict attention to the stable circular branch.

To determine whether the map \(r\mapsto z_{\max}(r)\) is monotonic, one must solve
\begin{equation}
\frac{d z_{\max}}{dr}=0.
\label{eq:dzdrzero}
\end{equation}
Introducing the auxiliary quantity
\begin{equation}
W(r)\equiv \sqrt{\frac{r f'(r)}{f(r)}},
\label{eq:Wdef}
\end{equation}
Eq. \eqref{eq:zmax_def} can be written as
\begin{equation}
1+z_{\max}(r)=\Xi\,\frac{\sqrt{2}+W(r)}{\sqrt{\mathcal{D}(r)}}.
\label{eq:zmax_compact}
\end{equation}
The extremum condition \eqref{eq:dzdrzero} is then equivalent to the exact
unsquared relation
\begin{equation}
2\mathcal{D}(r)\,\frac{dW}{dr}
=
\bigl(\sqrt{2}+W(r)\bigr)\,\mathcal{D}'(r),
\label{eq:extremum_compact}
\end{equation}
with
\begin{equation}
\mathcal{D}'(r)=f'(r)-r f''(r).
\end{equation}
Equation \eqref{eq:extremum_compact} is the cleanest form of the extremum
condition because it is algebraically exact and does not introduce spurious
roots.

By eliminating \(W'(r)\), Eq. \eqref{eq:extremum_compact} can also be written
entirely in terms of \(f(r)\) and its derivatives. \textcolor{black}{The possible loss of monotonicity is controlled by the exact condition}
\begin{equation}
\bigl(2f-r f'\bigr)\Bigl[f(f'+r f'')-r f'^2\Bigr]
=
f\bigl(f'-r f''\bigr)\Bigl[r f'+\sqrt{2r f f'}\Bigr],
\label{eq:extremum_exact}
\end{equation}
this is the exact condition for the appearance of a local extremum of
\(z_{\max}(r)\) in the physical domain. It should be stressed that any squared
version of Eq. \eqref{eq:extremum_exact} may generate spurious solutions and
must therefore be used, if at all, only as an intermediate root-finding device,
with all candidate roots checked against the original unsquared equation.
\textcolor{black}{The extremum condition must be supplemented by the physical-domain constraints
\begin{equation}\label{eqconditions}
f(r)>0,\qquad f'(r)>0,\qquad D(r)\equiv 2f-rf'>0,
\end{equation}
these inequalities ensure that the emitter is located outside the horizon,
that the circular orbit has positive angular momentum squared, and that the
spectroscopic observable is real and finite.} 
{\color{black}The global behavior of the local spectroscopic map depends on the parameters
\((\mu,\nu,q_m)\). A loss of monotonicity can occur only if
Eq.~\eqref{eq:extremum_exact} admits a solution satisfying the physical-domain
conditions \eqref{eqconditions}. We have checked the representative Fan--Wang
black-hole branches considered in this work and find no physical roots of
Eq.~\eqref{eq:extremum_exact} in the exterior circular-timelike domain
\(0\le q_m<q_{m,\rm ext}\). Consequently, for these representative branches the
local maximal-redshift branch remains globally monotonic and invertible. Any appearance of a
nonmonotonic branch in another member of the family would have to be established
by solving the unsquared extremum condition \eqref{eq:extremum_exact} together
with the constraints \eqref{eqconditions}.}

\textcolor{black}{The existence of a local extremum would imply a breakdown of the one-to-one correspondence between the maximal redshift and the orbital radius,
\begin{equation}
z_{\max}\longrightarrow {r_1,r_2},
\end{equation}
and would therefore signal a loss of invertibility of the spectroscopic map. Such a possibility is controlled by the exact extremum condition, Eq.~\eqref{eq:extremum_exact}. }
\textcolor{black}{Solving Eq.~(106) throughout the physical domain (107),
we find no physical roots for the representative
Fan-Wang branches considered in this work.
Consequently, the spectroscopic map remains globally
monotonic and invertible throughout the entire
black-hole sector $0 \le q_m < q_{\rm ext}$.}

{\color{black}It is nevertheless useful to compare this spectroscopic behavior with the extremal black-hole threshold. The extremal geometry is determined by \begin{equation} f(r_{\rm ext})=0, \qquad f'(r_{\rm ext})=0 . \end{equation} For the lapse function \eqref{eq:NED_f}, these two conditions give \begin{equation} r_{\rm ext}^{\nu} = (\mu-1)\,q_{m,\rm ext}^{\nu}, \label{eq:rext_relation} \end{equation} together with \begin{equation} q_{m,\rm ext} = \frac{2M\,(\mu-1)^{(\mu-1)/\nu}}{\mu^{\mu/\nu}}, \label{eq:qmext_corrected} \end{equation} and \begin{equation} r_{\rm ext} = 2M \left( \frac{\mu-1}{\mu} \right)^{\mu/\nu}. \label{eq:rext_corrected} \end{equation} At extremality the degenerate horizon may itself satisfy the circular-null condition and can therefore be regarded as a horizon null orbit. This horizon null orbit, however, should not be conflated with the exterior photon-sphere branch that bounds the exterior spectroscopic domain. The latter is obtained from \begin{equation} r f'(r)-2f(r)=0, \label{eq:outer_photon_sphere_NED} \end{equation} and, when present outside the horizon, determines the outer vacuum shadow impact parameter. Thus the extremal horizon structure and the exterior photon-sphere branch are distinct null circular structures and must be discussed separately.}

\begin{figure*}[!htbp]
    \centering
    \includegraphics[width=\linewidth]{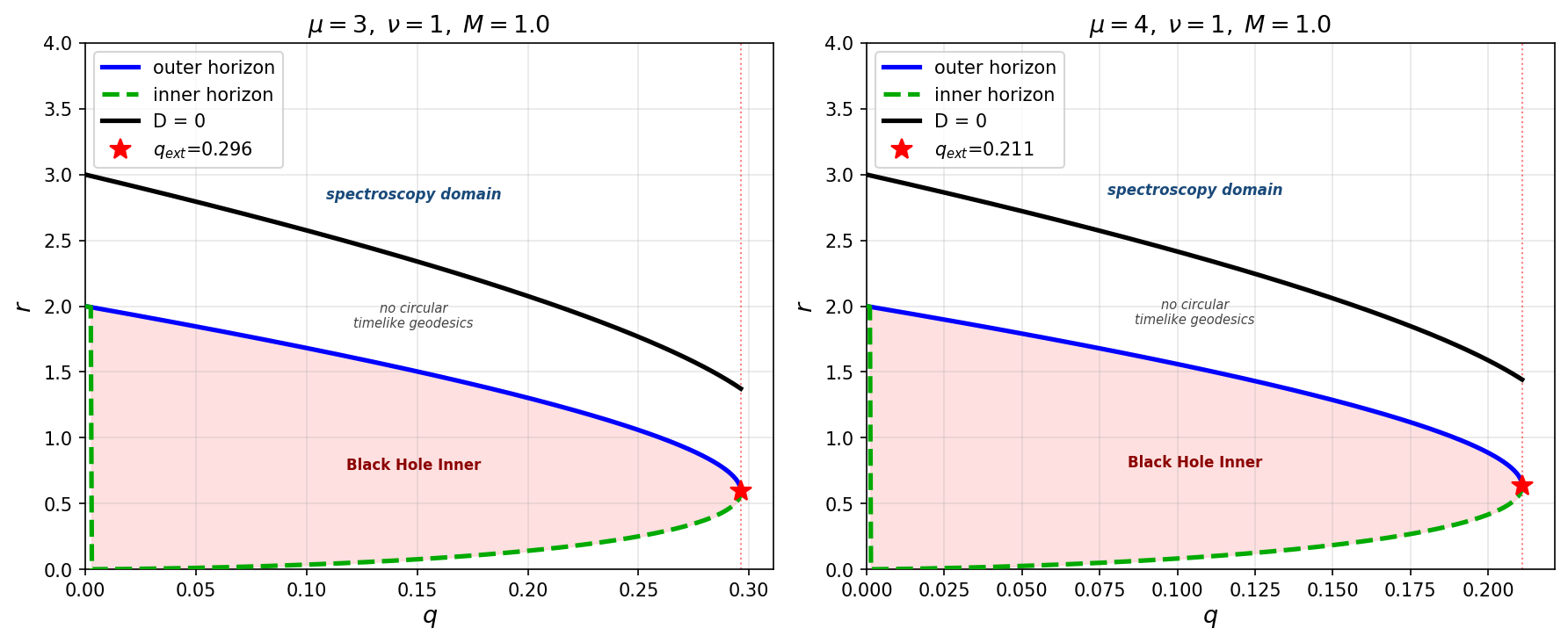}
    \caption{Phase structure of the Fan-Wang NED family in the $(q_m, r)$  plane for $\nu = 1$, $M = 1$, and two representative values of the  regularity parameter: $\mu = 3$ (left) and $\mu = 4$ (right). The solid blue curve marks the outer event horizon $r_+$, and the dashed green curve  marks the inner (Cauchy) horizon $r_-$, both obtained as roots of $f(r_h) = 0$. The shaded red region between the two horizons is the black hole interior. The solid black curve marks the photon sphere, defined by $D(r) \equiv 2f - rf^{\prime} = 0$. Between the outer horizon and the photon sphere, $D(r) < 0$ and no circular timelike geodesics exist; this region is therefore inaccessible to orbital spectroscopy. The region above the photon sphere, labeled \textit{spectroscopy domain}, is where $D(r) > 0$, circular timelike geodesics exist, and all frequency-shift observables $(R, S, T)$ of this paper are defined. At the extremal configuration the two horizons merge into a degenerate horizon (red star). This degenerate horizon may also satisfy the circular-null condition and can be regarded as a horizon null orbit, but it should not be confused with the outer photon-sphere branch that bounds the exterior spectroscopic domain. For $q_m > q_{\rm ext}$ (to the right of the vertical dotted red line) no horizon exists and the geometry does not describe a black hole; the spectroscopic framework of  this paper does not apply in that regime.}
    \label{fig:phase_structure}
\end{figure*}

\begin{figure*}[htp!]
\centering
\includegraphics[width=0.92\textwidth]{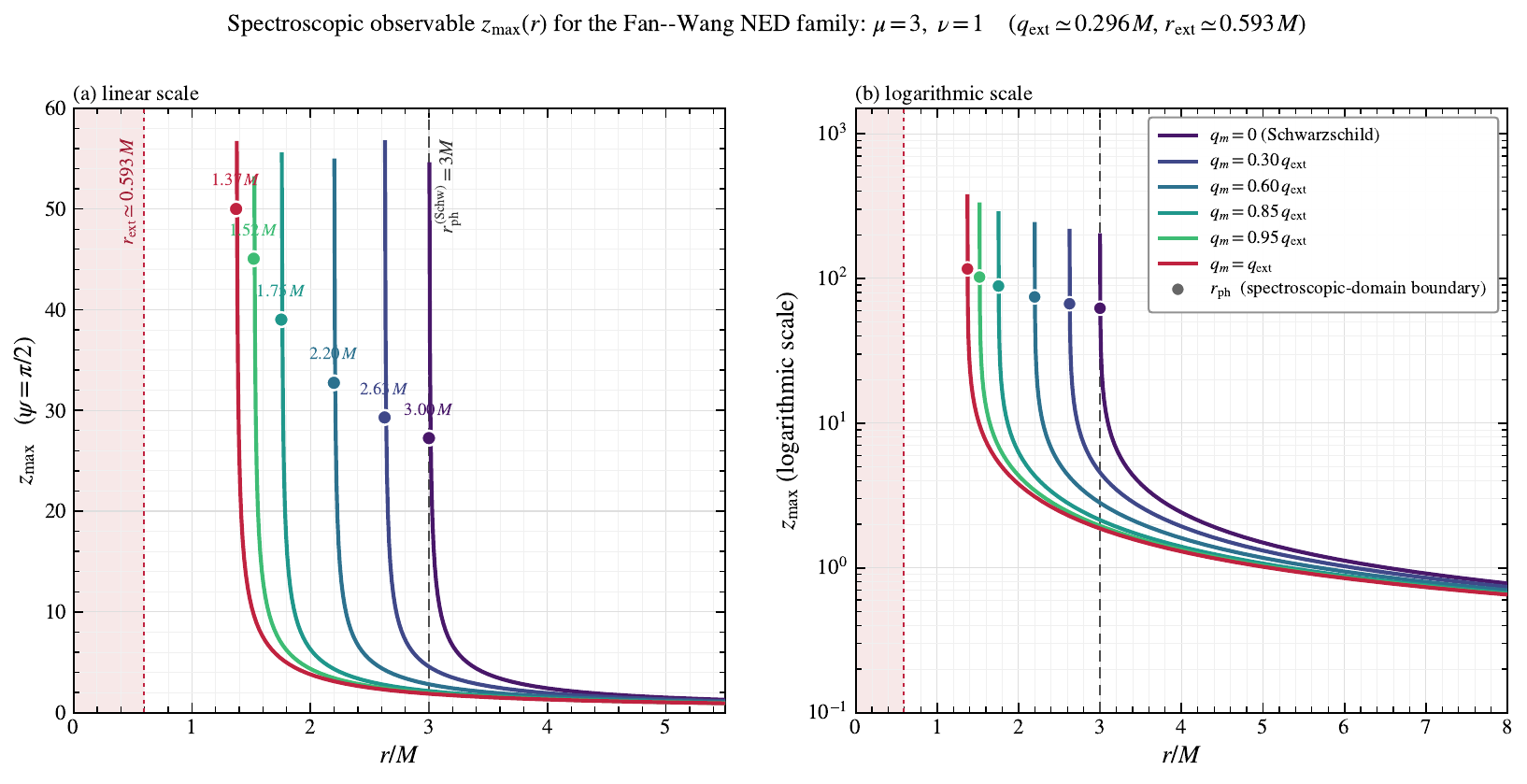}
\caption{Spectroscopic observable $z_{\max}(r)$ as a function of the emitter
radius $r/M$ for the Fan-Wang NED family (Eq. \eqref{eq:NED_redshift_explicit},
$\psi=\pi/2$), with $M=1$, $\mu=3$, $\nu=1$ (Maxwellian subfamily), and six representative values of the magnetic charge parameter $q_m$. Each colored curve starts at the spectroscopic domain boundary $r_{\rm ph}$ (filled circle), defined by $D(r)\equiv 2f-rf^{\prime}=0$, beyond which circular timelike geodesics exist. The vertical gray dashed line marks the Schwarzschild photon sphere at $r_{\rm ph}=3M$. As $q_m$ increases toward the extremal value $q_{\rm ext}$, the photon sphere shifts to smaller radii and $z_{\max}(r_{\rm ph}^{+})$ diverges, \textcolor{black}{reflecting the approach to the geometric boundary
$D(r)=0$ of the spectroscopic domain.} 
  The vertical red dashed line marks $r_{\rm ext}$, where the two horizons and the photon sphere merge at the extremal configuration. 
 \textcolor{black}{For $q_m\ll q_{\rm ext}$ the observable is strictly monotonically decreasing and closely follows the Schwarzschild profile. As $q_m$ increases, the shape of the curve is modified by the nonlinear electrodynamic corrections, but the observable remains monotonic throughout the physical black-hole sector. Consequently, the inverse map $z_{\max}\to r$ remains single-valued and globally invertible.}
}
\label{fig:zmax_NED_all}
\end{figure*}

\textcolor{black}{No physical spectroscopic degeneracy is found in the black-hole sector. Instead, the relevant geometric threshold is associated with the boundary of the spectroscopic domain and its approach to the extremal configuration. Figures~\ref{fig:phase_structure} and \ref{fig:zmax_NED_all} together provide a complete picture of this spectroscopic structure in the Fan-Wang NED family.
}

Figure~\ref{fig:phase_structure} establishes the global phase structure in the $(q_m, r)$ plane: the spectroscopic domain $D(r)>0$ lies above the photon-sphere curve, the black-hole interior is bounded by the two horizon branches, and all
three curves merge at the extremal point $(q_{\rm ext}, r_{\rm ext})$.
Figure~\ref{fig:zmax_NED_all} then resolves the internal structure of that domain by
displaying $z_{\max}(r)$ for six representative values of $q_m$: for
$q_m\ll q_{\rm ext}$ the observable is strictly monotonically decreasing and
indistinguishable in shape from the Schwarzschild curve, so the inverse map
$z_{\max}\to r$ is single-valued and unambiguous; as $q_m$ increases toward
$q_{\rm ext}$, the spectroscopic domain boundary $r_{\rm ph}$ (filled circles)
migrates toward $r_{\rm ext}$ along the photon-sphere locus visible in
Fig.~\ref{fig:phase_structure}, and the value $z_{\max}(r_{\rm ph}^{+})$ diverges,
signaling the geometric breakdown of the spectroscopic description at the photon
sphere where $D(r)\to 0$.
The two figures are therefore complementary: Fig.~\ref{fig:phase_structure} identifies
where in parameter space the spectroscopic domain exists and
when it collapses at extremality, while Fig.~\ref{fig:zmax_NED_all}
quantifies how the observable $z_{\max}(r)$ evolves within that domain and  demonstrates the global invertibility of the spectroscopic map.

\begin{figure}[!htbp]
    \centering
    \includegraphics[width=0.50\textwidth]{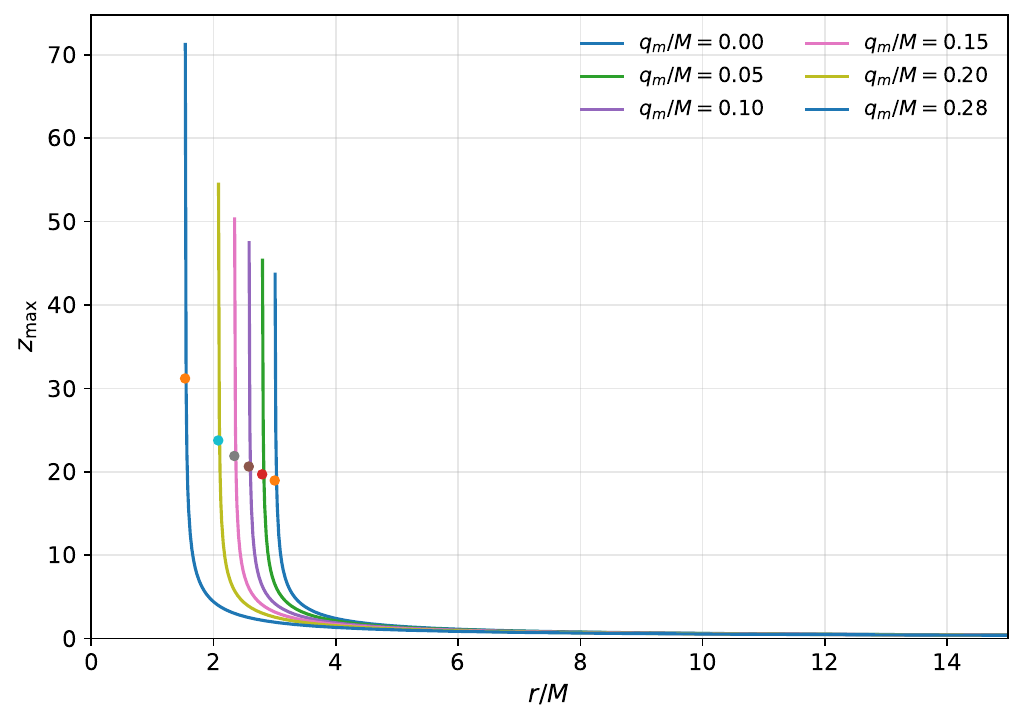}
    \caption{Maximal redshift \(z_{\max}(r)\) for the Fan-Wang family with
    \(\mu=3\), \(\nu=1\), and several representative values of the magnetic charge
    \(q_m/M\). Each curve begins just above the corresponding photon-sphere
    radius. As the charge increases, the lower bound of the spectroscopic domain
    shifts inward and the redshift grows more sharply near that boundary.}
    \label{fig:ned_zmax_curves}
\end{figure}

Figure~\ref{fig:ned_zmax_curves} shows that, for small and moderate charge, the
maximal redshift remains qualitatively close to the Schwarzschild behavior and
decreases monotonically with radius. As \(q_m/M\) increases, the lower edge of
the admissible domain moves inward and the observable becomes increasingly steep
near the photon-sphere boundary. This confirms that the strongest deviations from
the Schwarzschild profile arise in the high-charge, near-extremal sector.

\section{Example 2: the Simpson-Visser black-bounce spacetime}
\label{sec:SVexample}

As a second explicit application of the general framework, we consider the
Simpson-Visser black-bounce geometry, which furnishes a particularly simple and
instructive example of a spacetime interpolating between a regular black hole and
a traversable wormhole. In the form introduced in Ref.~\cite{Simpson:2018tsi},
the line element reads
\begin{eqnarray}
ds^{2}= -\left(1-\frac{2M}{\sqrt{r^{2}+b_0^{2}}}\right)dt^{2}\notag 
+ \left(1-\frac{2M}{\sqrt{r^{2}+b_0^{2}}}\right)^{-1}dr^{2}
&\quad \\ + (r^{2}+b_0^{2})\, (d\theta^2 + \sin^2\theta\, d\varphi^2),
\label{eq_SV_metric}
\end{eqnarray}
where \(M\) denotes the mass parameter and \(b_0\) is the bounce parameter.
The latter regularizes the Schwarzschild singularity and controls the global
structure of the spacetime. In the limit \(b_0\to0\), the geometry reduces
smoothly to Schwarzschild, whereas sufficiently large \(b_0\) removes the event
horizon and yields a two-way traversable wormhole.

It is useful to introduce the effective areal radius
\begin{equation}
R(r)\equiv \sqrt{r^{2}+b_0^{2}},
\label{eq:SV_Rdef}
\end{equation}
in terms of which the metric takes the compact form
\begin{equation}
ds^2=-f(r)\,dt^2+\frac{dr^2}{f(r)}+R(r)^2\,d\Omega^2,
\label{eq:SV_metric_compact}
\end{equation}
with
\begin{equation}
f(r)=1-\frac{2M}{R(r)}.
\label{eq:SV_f}
\end{equation}
Comparison with the general static, spherically symmetric ansatz shows that the
metric functions are identified as
\begin{equation}
\alpha(r)=f(r),\qquad
\beta(r)=R(r)^2=r^2+b_0^2,\qquad
\gamma(r)=\frac{1}{f(r)}.
\label{eq:SV_identification}
\end{equation}
Therefore, the exact formulas derived in the previous sections can be applied to
this geometry without further modification.

The horizon structure is determined by the zeros of the lapse function,
$f(r_h)=0.$
Using Eq. \eqref{eq:SV_f}, this condition implies $R(r_h)=2M,$
or equivalently
\begin{equation}
r_h^2=4M^2-b_0^2.
\label{eq:SV_horizon}
\end{equation}
It follows immediately that horizons exist only for
\begin{equation}
b_0\le 2M.
\end{equation}
For \(b_0<2M\), the spacetime describes a regular black hole with horizons on the
two branches of the extended radial coordinate. At the critical value
\begin{equation}
b_0=2M,
\end{equation}
the horizon coincides with the bounce surface \(r=0\). By contrast, for
\begin{equation}
b_0>2M,
\end{equation}
no horizon is present and the geometry describes a traversable wormhole with a
throat located at
\begin{equation}
r=0,
\qquad
R(0)=b_0.
\end{equation}

The first two derivatives of the lapse function are
\begin{equation}
f'(r)=\frac{2Mr}{(r^2+b_0^2)^{3/2}}
=\frac{2Mr}{R(r)^3},
\label{eq:SV_fprime}
\end{equation}
and
\begin{equation}
f''(r)=\frac{2M(b_0^2-2r^2)}{(r^2+b_0^2)^{5/2}}
=\frac{2M(b_0^2-2r^2)}{R(r)^5}.
\label{eq:SV_fdoubleprime}
\end{equation}
These expressions will enter directly into the circular-orbit and photon-sphere
conditions below.

We next specialize the general timelike-orbit formulas to the Simpson-Visser
metric. The denominator that controls the circular geodesics is
\begin{equation}
\alpha\beta'-\beta\alpha'
=
2r\,f(r)-R(r)^2 f'(r).
\end{equation}
Substituting Eqs. \eqref{eq:SV_Rdef} and \eqref{eq:SV_fprime} yields
\begin{equation}
\alpha\beta'-\beta\alpha'
=
2r\left(1-\frac{3M}{R(r)}\right).
\label{eq:SV_D}
\end{equation}
This combination is the Simpson-Visser counterpart of the standard
Schwarzschild factor \(2r(1-3M/r)\), with the areal radius \(R(r)\) replacing
the coordinate radius.

The energy and angular momentum per unit rest mass of a massive particle on a
circular orbit therefore become
\begin{equation}
E^2=
\frac{f(r)^2}{1-\dfrac{3M}{R(r)}},
\label{eq:SV_E2}
\end{equation}
and
\begin{equation}
L^2=
\frac{M\,R(r)}{1-\dfrac{3M}{R(r)}}
=
\frac{M\,R(r)^2}{R(r)-3M}.
\label{eq:SV_L2}
\end{equation}
Likewise, the emitter four-velocity components reduce to
\begin{equation}
u^t=
\frac{1}{\sqrt{1-\dfrac{3M}{R(r)}}},
\label{eq:SV_ut}
\end{equation}
and
\begin{equation}
u^\varphi=
\sqrt{\frac{M}{R(r)^2\,[R(r)-3M]}}.
\label{eq:SV_uphi}
\end{equation}

These formulas show that circular timelike motion is controlled by the condition
\begin{equation}
R(r)>3M.
\label{eq:SV_timelike_condition}
\end{equation}
In the Schwarzschild limit \(b_0\to0\), this reduces to the familiar relation
\(r>3M\). In the black-bounce geometry, however, it is the areal radius rather
than the coordinate \(r\) that controls the location of the circular-orbit
barrier.

We now turn to the null sector. For photons emitted at a local angle \(\psi\),
the general impact-parameter formula gives
\begin{equation}
b(r,\psi)=
-\frac{R(r)}{\sqrt{f(r)}}\sin\psi .
\label{eq:SV_b}
\end{equation}
Hence, for a static observer at infinity, the observed vacuum frequency shift is
\begin{equation}
1+z=u^t_{\rm e}-b\,u^\varphi_{\rm e}.
\label{eq:SV_redshift}
\end{equation}
Using Eqs. \eqref{eq:SV_ut}, \eqref{eq:SV_uphi}, and \eqref{eq:SV_b}, one obtains
\begin{equation}
1+z=
\frac{1}{\sqrt{1-\dfrac{3M}{R(r_{\rm e})}}}
+
\sin\psi\,
\sqrt{
\frac{M}
{R(r_{\rm e})\,f(r_{\rm e})
\left[1-\dfrac{3M}{R(r_{\rm e})}\right]}
}.
\label{eq:SV_redshift_explicit}
\end{equation}
If the peculiar motion of the source relative to the detector is included, then
\begin{equation}
\Xi\equiv \sqrt{\frac{1+\upsilon_0}{1-\upsilon_0}},
\end{equation}
and the total frequency shift is
\begin{equation}
1+z_{\rm tot}=\Xi\,(1+z).
\label{eq:SV_ztot}
\end{equation}

As in the general formalism, the distinguished emission directions are
\begin{equation}
\psi=0,\qquad \psi=\pm\pi/2.
\end{equation}
The case \(\psi=0\) corresponds to \textcolor{black}{radial emission shift} and isolates the
gravitational redshift, whereas \(\psi=\pm\pi/2\) yield the maximal redshift and
maximal blueshift, respectively.

The null critical structure is slightly richer than in Schwarzschild because
\(r=0\) is no longer a curvature singularity but a regular bounce surface. The
general photon-sphere condition
\begin{equation}
\frac{d}{dr}\left(\frac{\beta}{\alpha}\right)=0
\end{equation}
now becomes
\begin{equation}
\frac{d}{dr}\left(\frac{R(r)^2}{f(r)}\right)=0.
\label{eq:SV_phcond1}
\end{equation}
For \(r\neq0\), this reduces to the standard outer null-orbit condition
\begin{equation}
R(r_{\rm ph})=3M,
\end{equation}
from which one finds
\begin{equation}
r_{\rm ph}=\sqrt{9M^2-b_0^2},
\qquad
(b_0\le 3M).
\label{eq:SV_rph}
\end{equation}
The corresponding shadow impact parameter is therefore
\begin{equation}
b_{\rm sh}
=
\frac{R(r_{\rm ph})}{\sqrt{f(r_{\rm ph})}}
=
3\sqrt{3}\,M,
\qquad
(b_0\le 3M).
\label{eq:SV_shadow}
\end{equation}
Thus, the outer photon-sphere branch reproduces the Schwarzschild shadow radius
when expressed in terms of the mass parameter \(M\), even though the interior
geometry is completely regularized.

At the same time, one should keep in mind that the presence of the bounce surface
at \(r=0\) can enrich the full null critical structure, especially in the
wormhole regime. Accordingly, Eq. \eqref{eq:SV_rph} should be interpreted as the
condition for the standard outer photon sphere, while a complete classification
of all null critical curves may depend on the global causal structure and should
be analyzed separately when needed.

The Simpson-Visser geometry also fits naturally into the perturbative framework
developed in Sec.~\ref{sec:perturbative}. For large \(r\), the effective areal
radius has the asymptotic expansion
\begin{equation}
R(r)=r\left(1+\frac{b_0^2}{2r^2}-\frac{b_0^4}{8r^4}+\mathcal{O}(r^{-6})\right),
\end{equation}
which implies
\begin{equation}
\frac{1}{R(r)}
=
\frac{1}{r}
\left(1-\frac{b_0^2}{2r^2}+\frac{3b_0^4}{8r^4}+\mathcal{O}(r^{-6})\right).
\end{equation}
Accordingly, the lapse function admits the expansion
\begin{equation}
f(r)
=
1-\frac{2M}{r}
+\frac{M b_0^2}{r^3}
-\frac{3M b_0^4}{4r^5}
+\mathcal{O}(r^{-7}),
\label{eq:SV_large_r}
\end{equation}
which makes explicit how the bounce parameter deforms the Schwarzschild
background.

A natural dimensionless deformation parameter is therefore
\begin{equation}
\delta\sim \frac{b_0^2}{M^2},
\end{equation}
so that the metric functions may be organized as
\begin{equation}
\alpha(r,\delta)=f(r),\qquad
\beta(r,\delta)=r^2+b_0^2,\qquad
\gamma(r,\delta)=\frac{1}{f(r)}.
\end{equation}
In this form, the Simpson-Visser spacetime provides a clean example in which the
deviations from Schwarzschild enter the spectroscopy at even powers of the
bounce parameter, as expected from the symmetry \(r\leftrightarrow -r\).

The main significance of this example is that it shows how the general
spectroscopic framework applies not only to regular black holes sourced by
nonlinear electrodynamics, but also to black-bounce and wormhole geometries in
which the central singularity is replaced by a smooth throat-like structure.
This significantly broadens the range of compact-object models that can be
analyzed within a common spectroscopic language.

{\color{black}There is one additional subtlety in the Simpson--Visser geometry. The discussion above focuses on the standard outer photon-sphere branch. However, in the wormhole sector the throat is also part of the null critical structure. Writing \begin{equation} R(r)=\sqrt{r^2+b_0^2}, \qquad f(r)=1-\frac{2M}{R(r)}, \end{equation} one has \begin{equation} \frac{\beta(r)}{\alpha(r)} = \frac{R(r)^2}{f(r)} = \frac{R(r)^3}{R(r)-2M}. \label{eq:SV_h2} \end{equation} The vacuum critical condition gives \begin{equation} \frac{d}{dr} \left( \frac{\beta}{\alpha} \right) = \frac{2rR(R-3M)}{(R-2M)^2} =0 . \label{eq:SV_critical_full} \end{equation} Thus there are two kinds of critical structures: the usual outer branches \begin{equation} R=3M, \qquad r_{\rm ph}^{\pm} = \pm\sqrt{9M^2-b_0^2}, \qquad b_0\le 3M, \end{equation} and the throat branch \(r=0\). In the wormhole interval \(2M<b_0<3M\), the outer photon spheres coexist with a throat antiphoton sphere. At \(b_0=3M\), the outer branches merge with the throat. For \(b_0>3M\), the throat branch is the remaining relevant null-critical structure. In the present work we focus on the standard outer branch when comparing spectroscopy with the usual shadow construction, while noting that a complete lensing and shadow classification in the wormhole sector must also include the throat photon/antiphoton structure.}

\begin{figure}[!htbp]
    \centering
    \includegraphics[width=0.50\textwidth]{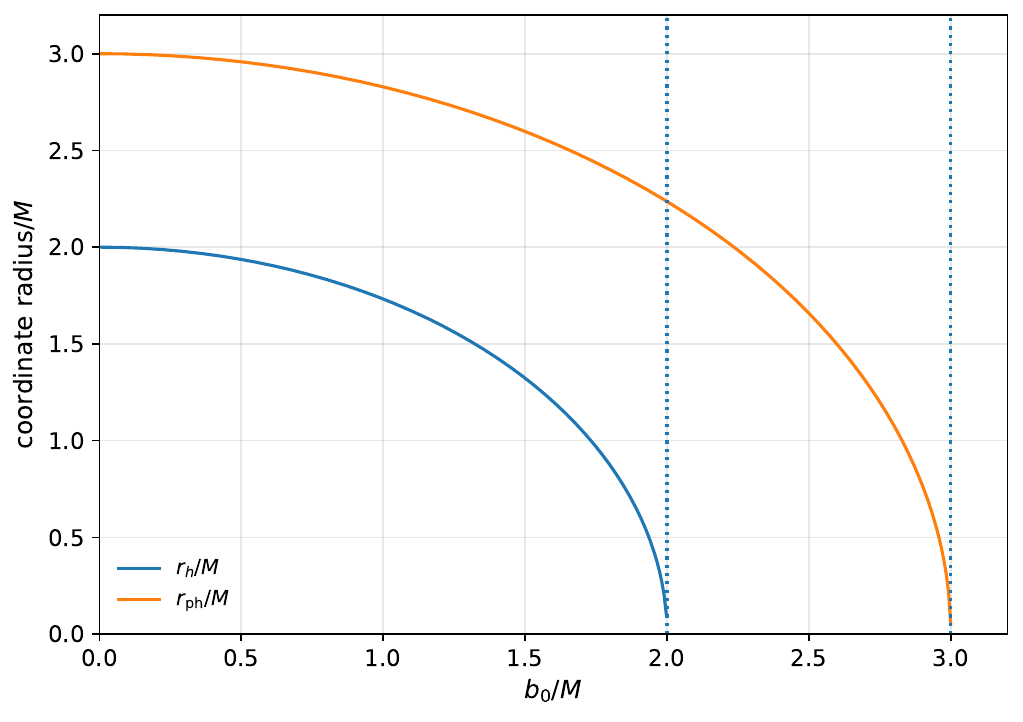}
\caption{{\color{black}Coordinate radii of the horizon, the standard outer photon sphere, and the throat critical structure in the Simpson--Visser geometry as functions of the bounce parameter \(b_0/M\). Horizons exist only for \(b_0\le 2M\), while the outer photon-sphere branch persists up to \(b_0=3M\). The interval \(2M<b_0<3M\) is wormhole-like and contains both the outer photon-sphere branch and a throat antiphoton sphere. At \(b_0=3M\) the outer branch merges with the throat, while for \(b_0>3M\) the throat branch is the remaining relevant null-critical structure.}}
    \label{fig:sv_transition_radii}
\end{figure}

Figure~\ref{fig:sv_transition_radii} summarizes the geometric transition encoded
by the bounce parameter. The horizon disappears at \(b_0=2M\), whereas the outer
photon sphere survives up to \(b_0=3M\). Consequently, there exists a finite
interval \(2M<b_0<3M\) in which the geometry is already wormhole-like but still
supports an outer photon sphere and therefore retains a nontrivial optical
structure relevant for spectroscopy and shadow formation.

\begin{figure}[!htbp]
    \centering
    \includegraphics[width=0.50\textwidth]{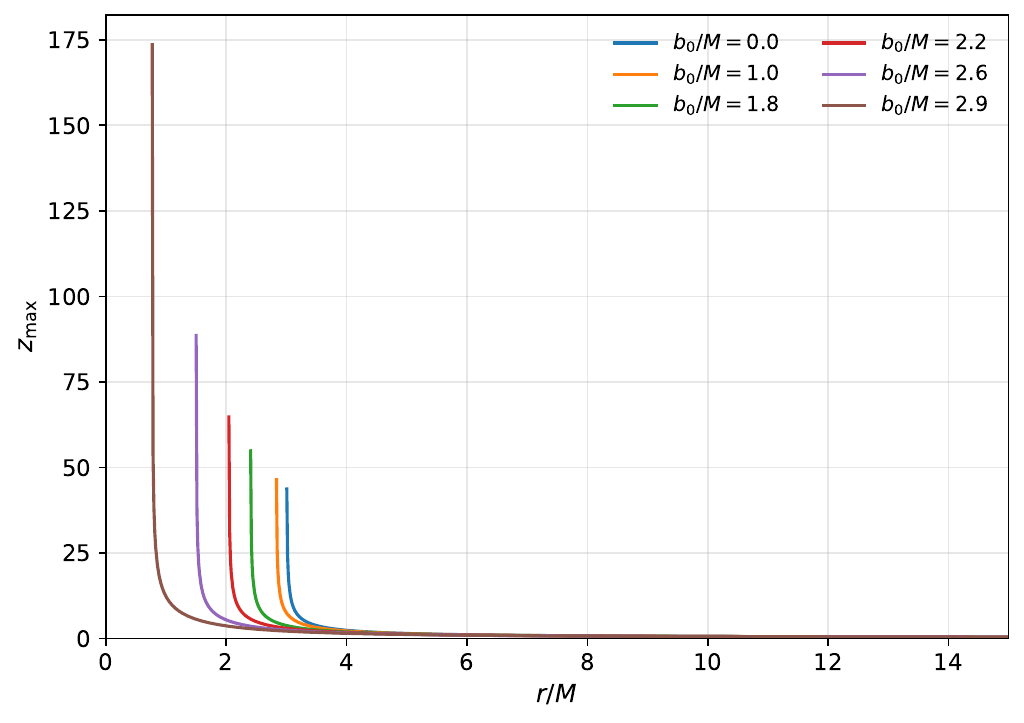}
 \caption{{\color{black}Local maximal-redshift branch \(z_{\max}(r)\) for the Simpson--Visser black-bounce spacetime for several representative values of the bounce parameter \(b_0/M\). The curves interpolate smoothly between the Schwarzschild case and the black-bounce/wormhole regime. The displayed domain is the circular-timelike domain and may include unstable circular geodesics; an astrophysical disk interpretation requires an additional restriction to the stable branch \(V_{\rm eff}''(r)>0\). The plotted local branch should not be interpreted as an observed line-profile extremum without solving the emitter--observer ray connection.}}
    \label{fig:sv_zmax_curves}
\end{figure}

Figure~\ref{fig:sv_zmax_curves} shows that the local maximal-redshift branch evolves smoothly
from the Schwarzschild case into the black-bounce and wormhole sectors. The main
effect of increasing \(b_0/M\) is to modify the lower edge of the spectroscopic
domain and to reshape the strong-field part of the curve, while preserving the
overall large-radius falloff. This behavior indicates that the Simpson-Visser
geometry provides a particularly clean setting in which regular black-hole and
wormhole spectroscopy can be studied within a single continuous family of
spacetimes.

\section{Example 3: the Fisher-Janis-Newman-Winicour-Wyman spacetime}
\label{sec:FJNWexample}
As a third explicit application of the general framework, we consider the Fisher-Janis-Newman-Winicour-Wyman spacetime, which is a static, spherically symmetric solution of Einstein gravity coupled to a massless scalar field \cite{Janis:1968zz,Wyman:1981bd,Virbhadra:1997ie,Sau:2020xau}. This geometry is especially useful in the present context because it probes a genuinely more general sector than the examples considered earlier: not only the temporal and radial metric functions but also the angular sector are nontrivially deformed. In this sense, it provides a sharper test of the model-independent formalism developed in Secs. \ref{sec:formalism} and \ref{sec:perturbative}.

We write the line element in the form
\begin{equation}
ds^{2}
=
-f(r)^{\nu} dt^{2}
+
f(r)^{-\nu} dr^{2}
+
r^{2} f(r)^{1-\nu}
\left(
d\theta^{2}
+
\sin^{2}\theta\, d\varphi^{2}
\right),
\label{VII.1}
\end{equation}
where
\begin{equation}
f(r)=1-\frac{r_{0}}{r},
\qquad
r_{0}=\frac{2M}{\nu},
\qquad
0<\nu\leq 1.
\label{VII.2}
\end{equation}
Here $M$ is the ADM mass, while the parameter $\nu$ measures the departure from Schwarzschild geometry and therefore the strength of the scalar-field deformation. In the limit $\nu\to 1$, one has $r_{0}\to 2M$ and Eq. \eqref{VII.1} reduces exactly to Schwarzschild. By contrast, for $0<\nu<1$, the surface $r=r_{0}$ is not an event horizon but a curvature singularity, so the spacetime describes a scalar-hairy naked singularity. The physically accessible exterior region is therefore
\begin{equation}
r>r_{0}.
\label{VII.3}
\end{equation}

Comparison with the general static, spherically symmetric ansatz shows that the metric functions are
\begin{equation}
\alpha(r)=f(r)^{\nu},
\qquad
\beta(r)=r^{2} f(r)^{1-\nu},
\qquad
\gamma(r)=f(r)^{-\nu}.
\label{VII.4}
\end{equation}
Unlike the Schwarzschild and Fan-Wang cases, the coordinate $r$ is not the areal radius. Instead, the area of the symmetry two-spheres is $4\pi \beta(r)$, so that the corresponding areal radius is $\sqrt{\beta(r)}=r\, f(r)^{(1-\nu)/2}$. This is precisely the feature that makes the FJNW/Wyman solution a nontrivial application of the unified formalism.

The radial derivative of the auxiliary function $f(r)$ is
\begin{equation}
f'(r)=\frac{r_{0}}{r^{2}},
\label{VII.5}
\end{equation}
from which we obtain
\begin{eqnarray}
\alpha'(r)=\nu \frac{r_{0}}{r^{2}} f(r)^{\nu-1},
\qquad &\\
\beta'(r)=2r\, f(r)^{1-\nu}+(1-\nu) r_{0} f(r)^{-\nu}.
\label{VII.6}
\end{eqnarray}
Substituting these expressions into the universal kernel controlling circular timelike motion, we find
\begin{equation}
\alpha \beta'-\beta \alpha'
=
2r-(1+2\nu) r_{0}.
\label{VII.7}
\end{equation}
This result is noteworthy because the same combination also controls the null critical structure. Hence, in this geometry as well, orbital spectroscopy and photon-sphere physics are governed by one and the same geometric object.

We now specialize the timelike circular geodesics. Using the general formulas of Sec.~II, the conserved energy and angular momentum per unit rest mass are
\begin{equation}
E^{2}
=
f(r)^{\nu}
\frac{2r-(1+\nu) r_{0}}{2r-(1+2\nu) r_{0}},
\label{VII.8}
\end{equation}
and
\begin{equation}
L^{2}
=
\frac{\nu r_{0}\, r^{2} f(r)^{1-\nu}}{2r-(1+2\nu) r_{0}}.
\label{VII.9}
\end{equation}
The corresponding nonvanishing components of the emitter four-velocity are
\begin{equation}
u^{t}
=
f(r)^{-\nu/2}
\sqrt{\frac{2r-(1+\nu) r_{0}}{2r-(1+2\nu) r_{0}}},
\label{VII.10}
\end{equation}
and
\begin{equation}
u^{\varphi}
=
f(r)^{(\nu-1)/2}
\sqrt{\frac{\nu r_{0}}{r^{2}\left[2r-(1+2\nu) r_{0}\right]}}.
\label{VII.11}
\end{equation}
Therefore, circular timelike emitters exist only when
\begin{equation}
2r-(1+2\nu) r_{0}>0,
\label{VII.12}
\end{equation}
or equivalently,
\begin{equation}
r>\frac{1+2\nu}{2}\, r_{0}.
\label{VII.13}
\end{equation}
For $\nu>1/2$, this lower bound lies outside the singular surface and defines an exterior circular-orbit barrier. For $0<\nu\leq 1/2$, the bound lies at or inside $r=r_{0}$, so the whole exterior region $r>r_{0}$ is, in principle, accessible to circular timelike motion. This is one of the main qualitative differences between the FJNW/Wyman geometry and the black-hole examples discussed previously.

We next turn to the null sector. For photons emitted at the local angle \(\psi\), measured in the static
orthonormal frame at the source, the impact parameter becomes
\begin{equation}
b(r,\psi)
=
-r\, f(r)^{(1-2\nu)/2}\sin\psi .
\label{VII.14}
\end{equation}
Accordingly, for a static observer at infinity, the observed vacuum frequency shift takes the exact form
\begin{eqnarray}
1+z
=
f(r_{\rm e})^{-\nu/2}
\sqrt{\frac{2r_{\rm e}-(1+\nu) r_{0}}
{2r_{\rm e}-(1+2\nu) r_{0}}}
\\+ \notag
\sin\psi\,
f(r_{\rm e})^{-\nu/2}
\sqrt{\frac{\nu r_{0}}
{2r_{\rm e}-(1+2\nu) r_{0}}}.
\label{VII.15}
\end{eqnarray}
If the source possesses a radial peculiar motion relative to the detector, then
\begin{equation}
\Xi=\sqrt{\frac{1+\upsilon_0}{1-\upsilon_0}},
\label{VII.16}
\end{equation}
and the total observed shift is
\begin{equation}
1+z_{\rm tot}=\Xi(1+z).
\label{VII.17}
\end{equation}
The three distinguished emission directions remain \(\psi=0\) and
\(\psi=\pm\pi/2\). For radial emission, we obtain the radial-emission branch contribution
\begin{equation}
1+z_{\rm g}
=
\Xi\, f(r_{\rm e})^{-\nu/2}
\sqrt{\frac{2r_{\rm e}-(1+\nu) r_{0}}{2r_{\rm e}-(1+2\nu) r_{0}}},
\label{VII.18}
\end{equation}
whereas tangential emission yields the extremal shifts
\begin{widetext}
\begin{eqnarray}
1+z_{\pm}=
\Xi\, f(r_{\rm e})^{-\nu/2}
\left[
\sqrt{\frac{2r_{\rm e}-(1+\nu) r_{0}}{2r_{\rm e}-(1+2\nu) r_{0}}}
\pm
\sqrt{\frac{\nu r_{0}}{2r_{\rm e}-(1+2\nu) r_{0}}}
\right].
\label{VII.19}
\end{eqnarray}
\end{widetext}
These formulas show explicitly how the scalar deformation parameter $\nu$ enters both the gravitational and kinematical parts of the signal. In particular, the strong-field enhancement is controlled by the same denominator that appeared in Eqs. \eqref{VII.8}-\eqref{VII.13}, which is why the spectroscopic signal becomes increasingly sensitive as the emitter approaches the circular-orbit boundary.

The null critical structure follows from the general photon-sphere condition. Since
\begin{equation}
\frac{\beta}{\alpha}=r^{2} f(r)^{1-2\nu},
\label{VII.20}
\end{equation}
the equation $d(\beta/\alpha)/dr=0$ yields
\begin{equation}
2r-(1+2\nu) r_{0}=0.
\label{VII.21}
\end{equation}
Thus, whenever the solution lies in the exterior region, the photon sphere is located at
\begin{equation}
r_{\rm ph}=\frac{1+2\nu}{2}\, r_{0}
=\frac{1+2\nu}{\nu}\, M.
\label{VII.22}
\end{equation}
This radius exists outside the singular surface only if
\begin{equation}
\nu>\frac{1}{2}.
\label{VII.23}
\end{equation}
Hence the FJNW/Wyman family exhibits a sharp threshold. For $\nu>1/2$, the spacetime possesses an exterior photon sphere and therefore supports the standard shadow construction. For $0<\nu\leq 1/2$, no exterior photon sphere exists, even though the frequency-shift observables remain well defined in the whole exterior region $r>r_{0}$. This distinction is physically important because it separates the regimes in which spectroscopy and shadow formation remain tightly linked from those in which spectroscopy survives without a standard unstable circular null orbit.

When $\nu>1/2$, the corresponding shadow impact parameter is
\begin{equation}
b_{\rm sh}
=
r_{\rm ph}\, f(r_{\rm ph})^{(1-2\nu)/2}
=
\frac{1+2\nu}{2}\, r_{0}
\left(
\frac{2\nu-1}{1+2\nu}
\right)^{(1-2\nu)/2}.
\label{VII.24}
\end{equation}
Equivalently,
\begin{equation}
b_{\rm sh}
=
\frac{1+2\nu}{2}\, r_{0}
\left(
\frac{1+2\nu}{2\nu-1}
\right)^{(2\nu-1)/2}.
\label{VII.25}
\end{equation}
In the Schwarzschild limit $\nu=1$ and $r_{0}=2M$, Eq. \eqref{VII.25} reduces to $b_{\rm sh}=3\sqrt{3}\,M$, as expected. This limiting-case check confirms that the present formulas are consistent with the standard black-hole result.

The plasma extension also follows immediately from the general formalism. For the power-law density profile introduced in Sec.~II, the refractive index becomes
\begin{equation}
n^{2}(r)=1-\frac{k}{r^{h}}\, f(r)^{\nu},
\label{VII.26}
\end{equation}
and the plasma-corrected impact parameter is
\begin{equation}
\hat b(r,\psi)
=
-n(r)\,r\, f(r)^{(1-2\nu)/2}\sin\psi .
\label{VII.27}
\end{equation}

{\color{black}As in the general discussion, this plasma-corrected impact parameter is used here only for the local spectroscopic frequency shift. The plasma-modified critical-ray condition would instead follow from extremizing \(b_{\rm pl}^2(r;\omega_0)=\beta n^2/\alpha\), and a complete plasma line-profile or shadow calculation would require solving the dispersive ray equation between the emitter and the observer.}

The total plasma-corrected frequency shift is therefore
\begin{widetext}
\begin{equation}
1+\hat z_{\rm tot}
=
\Xi
\left[
f(r_{\rm e})^{-\nu/2}
\sqrt{\frac{2r_{\rm e}-(1+\nu) r_{0}}
{2r_{\rm e}-(1+2\nu) r_{0}}}
+
n(r_{\rm e})\sin\psi\,
f(r_{\rm e})^{-\nu/2}
\sqrt{\frac{\nu r_{0}}
{2r_{\rm e}-(1+2\nu) r_{0}}}
\right].
\label{VII.28}
\end{equation}
\end{widetext}
As in the previous sections, the radial-emission branch remains unchanged by the plasma because the optical correction vanishes for \(\psi=0\). The dispersive medium therefore affects only the nonradial part of the signal.

It is also useful to display the large-radius behavior of the metric functions at fixed ADM mass $M$. Using $r_{0}=2M/\nu$, we obtain
\begin{equation}
\alpha(r)
=
1-\frac{2M}{r}
+
\frac{2(\nu-1)M^{2}}{\nu r^{2}}
+
\mathcal{O}(r^{-3}),
\label{VII.29}
\end{equation}
\begin{equation}
\beta(r)
=
r^{2}
-
\frac{2(1-\nu)M}{\nu}\, r
-
\frac{2(1-\nu)M^{2}}{\nu}
+
\mathcal{O}(r^{-1}),
\label{VII.30}
\end{equation}
and
\begin{equation}
\gamma(r)
=
1+\frac{2M}{r}
+
\frac{2(\nu+1)M^{2}}{\nu r^{2}}
+
\mathcal{O}(r^{-3}).
\label{VII.31}
\end{equation}
These expansions make clear that the FJNW/Wyman spacetime is asymptotically flat and approaches Schwarzschild at large radius, but they also show that the nontrivial angular sector survives already at subleading order. For this reason, a natural deformation parameter is
\begin{equation}
\delta \sim 1-\nu,
\label{VII.32}
\end{equation}
although, in this coordinate representation, the deformation is distributed simultaneously among $\alpha$, $\beta$, and $\gamma$. This is precisely why the spacetime furnishes a more demanding test of the perturbative and exact spectroscopic framework than examples with $\beta(r)=r^{2}$.

The main significance of the FJNW/Wyman example is therefore twofold. First, it shows that the unified spectroscopic formalism extends smoothly from regular black holes and black-bounce geometries to scalar-supported horizonless compact objects. Second, it reveals a sharp geometric threshold at $\nu=1/2$: above it, spectroscopy and shadow formation are controlled by the same exterior photon sphere, whereas below it, the spectroscopic observables remain well defined even though the standard shadow construction ceases to exist. This makes the FJNW/Wyman family a particularly instructive example for disentangling which strong-field observables genuinely require an external null critical orbit and which do not.

\begin{figure}[!htbp]
    \centering
    \includegraphics[width=0.95\linewidth]{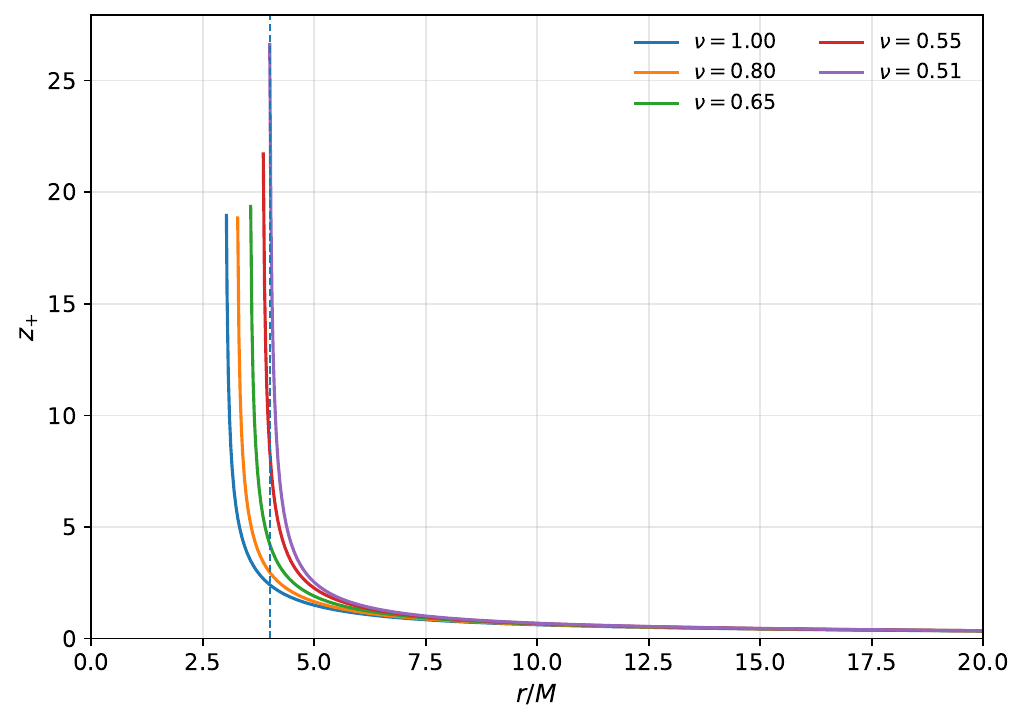}
    \caption{Local maximal redshift branch \(z_{+}(r)\) in the Fisher-Janis-Newman-Winicour-Wyman spacetime for several representative values of the scalar-deformation parameter \(\nu\), approaching the critical threshold \(\nu=1/2\) from above. Unlike the charged regular-black-hole examples, the FJNW geometry does not possess a magnetic-charge extremality bound. Instead, the relevant sharp transition occurs at \(\nu=1/2\), below which the exterior photon sphere disappears. The plot shows how the strong-field spectroscopic signal changes as the geometry approaches this threshold.}
    \label{fig:fjnw_zplus_near_threshold}
\end{figure}

Figure~\ref{fig:fjnw_zplus_near_threshold} shows how the local maximal redshift branch evolves
as the scalar-deformation parameter \(\nu\) moves away from the Schwarzschild
limit \(\nu=1\) and approaches the critical value \(\nu=1/2\). The strongest
deviations occur in the near-field region, where the denominator
\(2r-(1+2\nu)r_0\) becomes small and the spectroscopic signal is enhanced.
As \(\nu\to 1/2^{+}\), the geometry approaches the threshold at which the
exterior photon sphere disappears. The plot therefore illustrates the FJNW
analog of a near-extremal transition: not the approach to a charged horizon
merger, but the approach to the loss of the standard unstable circular null
orbit while the spectroscopic observables remain well defined.

\section{Conclusion}
\label{con}

In this paper we have developed a unified analytical framework for the
spectroscopy of photon frequency shifts in generic static, spherically symmetric
spacetimes. Starting from a general metric written in terms of the radial
functions \(\alpha(r,\delta)\), \(\beta(r,\delta)\), and \(\gamma(r,\delta)\),
we derived exact expressions for the conserved quantities of massive and null
probes, the conditions for circular timelike motion, the local-emission-angle
dependence of the photon impact parameter, and the corresponding redshift and
blueshift signals measured by distant observers.

{\color{black}A central outcome of the analysis is that, in vacuum, orbital spectroscopy and critical null structure can be described within one common geometric language. The same metric functions that determine the local redshift/blueshift branches of photons emitted by circularly orbiting sources also determine the photon sphere and the shadow impact parameter whenever an external null critical orbit exists. This provides a conceptual and phenomenological bridge between local spectroscopic observables and black-hole imaging observables, while also making clear where the connection can fail or require refinement, for example in horizonless spacetimes without an external photon sphere or in dispersive plasma environments.}

{\color{black}We further extended the local spectroscopic formalism to a nonmagnetized cold plasma. In this setting, the refractive index modifies the local impact parameter and therefore the nonradial frequency-shift branches, while the radial-emission branch is unchanged at the level of the local frequency shift. We also displayed the corresponding plasma-modified critical-ray condition. However, a full calculation of plasma-modified shadows, observed line profiles, or image-dependent extrema requires solving the dispersive ray problem between the emitter and the observer, which lies beyond the scope of the present analytic treatment.}

To make the framework suitable for a wide class of deformed compact objects, we
also constructed a perturbative expansion around Schwarzschild geometry up to
second order in a dimensionless deformation parameter \(\delta\). The resulting
hierarchy of corrections to the orbital energy, angular momentum, emitter
four-velocity, photon impact parameter, and vacuum and plasma redshift/blueshift
observables identifies the precise combinations of deformation functions that
enter the spectroscopy at each order. In this sense, the perturbative sector of
the paper provides a practical bridge between exact formalism and model-specific
phenomenology.

{\color{black}As explicit illustrations, we applied the formalism to regular black holes from nonlinear electrodynamics, to the Simpson--Visser black-bounce geometry, and to the Fisher--Janis--Newman--Winicour--Wyman spacetime. These examples show that the same local spectroscopic framework can be used across regular-black-hole, black-bounce, wormhole-like, and scalar-supported horizonless sectors. In the Fan--Wang nonlinear-electrodynamics family, the representative black-hole branches studied here show a globally monotonic local maximal-redshift branch in the exterior circular-timelike domain. In the Simpson--Visser case, the analysis also highlights the need to distinguish the standard outer photon-sphere branch from the throat photon/antiphoton structure in the wormhole sector. The Fisher--Janis--Newman--Winicour--Wyman example further shows that local spectroscopic observables may remain well defined even when the standard external photon sphere is absent.}

Several natural extensions suggest themselves. One important next step is a
joint spectroscopic-imaging analysis in which the observable triplets
\((R,S,T)\) and \((\hat R,\hat S,\hat T)\) are combined with photon-sphere or
shadow data in order to improve the reconstruction of the underlying spacetime
parameters. Another promising direction is the development of Bayesian parameter
estimation pipelines using Galactic-center and megamaser observations
\cite{Nucamendi:2021iws,VillalobosRamirez:2022,Villaraos:2022,Momennia:2024rapidity},
for which the present perturbative formulas may serve as efficient analytical
templates. More broadly, the framework developed here may be adapted to other
compact-object environments and used as a basis for systematic tests of
deviations from Schwarzschild geometry in strong gravitational fields.

\acknowledgments
A. \"O., and R. P. would like to acknowledge networking support of the COST Action CA21106 - COSMIC WISPers in the Dark Universe: Theory, astrophysics and experiments (CosmicWISPers), the COST Action CA22113 - Fundamental challenges in theoretical physics (THEORY-CHALLENGES), the COST Action CA21136 - Addressing observational tensions in cosmology with systematics and fundamental physics (CosmoVerse), the COST Action CA23130 - Bridging high and low energies in search of quantum gravity (BridgeQG), and the COST Action CA23115 - Relativistic Quantum Information (RQI) funded by COST (European Cooperation in Science and Technology). J. S. acknowledges the FONDECYT
grant N°1220065, Chile.

\appendix
\section{Derivation of the local impact-parameter formula}
\label{app:impact_derivation}

For completeness, we derive the impact-parameter relation used in the main text.
The static orthonormal tetrad adapted to the metric
\[
ds^2=-\alpha dt^2+\gamma dr^2+\beta(d\theta^2+\sin^2\theta d\varphi^2)
\]
is
\begin{equation}
e_{(t)}=\alpha^{-1/2}\partial_t,\qquad
e_{(r)}=\gamma^{-1/2}\partial_r,\qquad
e_{(\varphi)}=\beta^{-1/2}\partial_\varphi .
\end{equation}
For a photon in the equatorial plane, the tetrad components satisfy
\begin{equation}
k^{(t)}=\sqrt{\alpha}\,k^t=\frac{E_\gamma}{\sqrt{\alpha}},
\qquad
k^{(\varphi)}=\sqrt{\beta}\,k^\varphi=\frac{L_\gamma}{\sqrt{\beta}} .
\end{equation}
The local emission angle \(\psi\) measured by the static observer is defined by
\begin{equation}
\sin\psi=\frac{k^{(\varphi)}}{k^{(t)}} .
\end{equation}
Using \(b=L_\gamma/E_\gamma\), one obtains
\begin{equation}
\sin\psi
=
b\sqrt{\frac{\alpha}{\beta}},
\end{equation}
and therefore
\begin{equation}
b
=
-\sqrt{\frac{\beta}{\alpha}}\,\sin\psi ,
\end{equation}
where the minus sign fixes the convention used for the redshift/blueshift
branches in the main text.

\section{Perturbative coefficients}
\label{app:perturbative_details}

\subsection{Generic quotient and square-root expansions}

For later use, we employ
\begin{equation}
\frac{N_0+\delta N_1+\delta^2 N_2}{D_0+\delta D_1+\delta^2 D_2}
=
Q_0+\delta Q_1+\delta^2 Q_2+\mathcal{O}(\delta^3),
\end{equation}
where
\begin{align}
Q_0&=\frac{N_0}{D_0},\\
Q_1&=\frac{N_1D_0-N_0D_1}{D_0^2},\\
Q_2&=\frac{N_2D_0^2-N_1D_0D_1+N_0(D_1^2-D_0D_2)}{D_0^3},
\end{align}
and
\begin{eqnarray}
\sqrt{Q_0+\delta Q_1+\delta^2 Q_2}
=
\sqrt{Q_0}
+\delta\,\frac{Q_1}{2\sqrt{Q_0}}
\\+\delta^2\left(\frac{Q_2}{2\sqrt{Q_0}}-\frac{Q_1^2}{8Q_0^{3/2}}\right)
+\mathcal{O}(\delta^3). \notag
\end{eqnarray}

\subsection{Circular timelike geodesics}

For the energy expansion,
\begin{align}
\mathcal{N}^{(E)}_0&=\alpha_0^2\beta_0',\\
\mathcal{N}^{(E)}_1&=\alpha_0^2\beta_1'+2\alpha_0\alpha_1\beta_0',\\
\mathcal{N}^{(E)}_2&=\alpha_0^2\beta_2'+2\alpha_0\alpha_1\beta_1'
+\left(2\alpha_0\alpha_2+\alpha_1^2\right)\beta_0',
\end{align}
so that
\begin{align}
E_{(0)}^2&=\frac{\mathcal{N}^{(E)}_0}{\mathcal{D}_0},\\
E_{(1)}^2&=\frac{\mathcal{N}^{(E)}_1\mathcal{D}_0-\mathcal{N}^{(E)}_0\mathcal{D}_1}{\mathcal{D}_0^2},\\
E_{(2)}^2&=
\frac{\mathcal{N}^{(E)}_2\mathcal{D}_0^2-\mathcal{N}^{(E)}_1\mathcal{D}_0\mathcal{D}_1
+\mathcal{N}^{(E)}_0(\mathcal{D}_1^2-\mathcal{D}_0\mathcal{D}_2)}
{\mathcal{D}_0^3}.
\end{align}

For the angular momentum expansion,
\begin{align}
\mathcal{N}^{(L)}_0&=\beta_0^2\alpha_0',\\
\mathcal{N}^{(L)}_1&=\beta_0^2\alpha_1'+2\beta_0\beta_1\alpha_0',\\
\mathcal{N}^{(L)}_2&=\beta_0^2\alpha_2'+2\beta_0\beta_1\alpha_1'
+\left(2\beta_0\beta_2+\beta_1^2\right)\alpha_0',
\end{align}
so that
\begin{align}
L_{(0)}^2&=\frac{\mathcal{N}^{(L)}_0}{\mathcal{D}_0},\\
L_{(1)}^2&=\frac{\mathcal{N}^{(L)}_1\mathcal{D}_0-\mathcal{N}^{(L)}_0\mathcal{D}_1}{\mathcal{D}_0^2},\\
L_{(2)}^2&=
\frac{\mathcal{N}^{(L)}_2\mathcal{D}_0^2-\mathcal{N}^{(L)}_1\mathcal{D}_0\mathcal{D}_1
+\mathcal{N}^{(L)}_0(\mathcal{D}_1^2-\mathcal{D}_0\mathcal{D}_2)}
{\mathcal{D}_0^3}.
\end{align}

For the four-velocity components, define
\begin{align}
\mathcal{T}_0&=\frac{\beta_0'}{\mathcal{D}_0},\\
\mathcal{T}_1&=\frac{\beta_1'\mathcal{D}_0-\beta_0'\mathcal{D}_1}{\mathcal{D}_0^2},\\
\mathcal{T}_2&=\frac{\beta_2'\mathcal{D}_0^2-\beta_1'\mathcal{D}_0\mathcal{D}_1
+\beta_0'(\mathcal{D}_1^2-\mathcal{D}_0\mathcal{D}_2)}{\mathcal{D}_0^3},
\end{align}
and
\begin{align}
\Phi_0&=\frac{\alpha_0'}{\mathcal{D}_0},\\
\Phi_1&=\frac{\alpha_1'\mathcal{D}_0-\alpha_0'\mathcal{D}_1}{\mathcal{D}_0^2},\\
\Phi_2&=\frac{\alpha_2'\mathcal{D}_0^2-\alpha_1'\mathcal{D}_0\mathcal{D}_1
+\alpha_0'(\mathcal{D}_1^2-\mathcal{D}_0\mathcal{D}_2)}{\mathcal{D}_0^3}.
\end{align}
Then
\begin{align}
u^t_{(0)}&=\sqrt{\mathcal{T}_0},\\
u^t_{(1)}&=\frac{\mathcal{T}_1}{2\sqrt{\mathcal{T}_0}},\\
u^t_{(2)}&=\frac{\mathcal{T}_2}{2\sqrt{\mathcal{T}_0}}-\frac{\mathcal{T}_1^2}{8\mathcal{T}_0^{3/2}},
\end{align}
and
\begin{align}
u^\varphi_{(0)}&=\sqrt{\Phi_0},\\
u^\varphi_{(1)}&=\frac{\Phi_1}{2\sqrt{\Phi_0}},\\
u^\varphi_{(2)}&=\frac{\Phi_2}{2\sqrt{\Phi_0}}-\frac{\Phi_1^2}{8\Phi_0^{3/2}}.
\end{align}

\subsection{Photon impact parameter}

Writing
\begin{equation}
s\equiv \sin\psi ,
\end{equation}
the local-angle impact parameter is
\begin{equation}
b=-s\sqrt{\frac{\beta}{\alpha}} .
\end{equation}
We define
\begin{equation}
\mathcal{B}\equiv \frac{\beta}{\alpha}
=\mathcal{B}_0+\delta\mathcal{B}_1+\delta^2\mathcal{B}_2
+\mathcal{O}(\delta^3),
\end{equation}
where
\begin{align}
\mathcal{B}_0&=\frac{\beta_0}{\alpha_0},\\
\mathcal{B}_1&=\frac{\beta_1\alpha_0-\beta_0\alpha_1}{\alpha_0^2},\\
\mathcal{B}_2&=
\frac{\beta_2\alpha_0^2-\beta_1\alpha_0\alpha_1
+\beta_0(\alpha_1^2-\alpha_0\alpha_2)}{\alpha_0^3}.
\end{align}
Thus
\begin{align}
b_{(0)}&=-s\sqrt{\mathcal{B}_0},\\
b_{(1)}&=-s\,\frac{\mathcal{B}_1}{2\sqrt{\mathcal{B}_0}},\\
b_{(2)}&=-s\left(
\frac{\mathcal{B}_2}{2\sqrt{\mathcal{B}_0}}
-\frac{\mathcal{B}_1^2}{8\mathcal{B}_0^{3/2}}
\right).
\end{align}

\subsection{Redshift coefficients}

The vacuum redshift coefficients are
\begin{align}
\mathcal{Z}_0&=u^t_{(0)}-b_{(0)}u^\varphi_{(0)},\\
\mathcal{Z}_1&=u^t_{(1)}-b_{(0)}u^\varphi_{(1)}-b_{(1)}u^\varphi_{(0)},\\
\mathcal{Z}_2&=u^t_{(2)}-b_{(0)}u^\varphi_{(2)}-b_{(1)}u^\varphi_{(1)}-b_{(2)}u^\varphi_{(0)}.
\end{align}

The corresponding triplet coefficients are
\begin{align}
R_{(0)}&=\Xi\,\mathcal{Z}_0\big|_{\psi=\pi/2},\\
R_{(1)}&=\Xi\,\mathcal{Z}_1\big|_{\psi=\pi/2},\\
R_{(2)}&=\Xi\,\mathcal{Z}_2\big|_{\psi=\pi/2},\\
S_{(0)}&=\Xi\,\mathcal{Z}_0\big|_{\psi=-\pi/2},\\
S_{(1)}&=\Xi\,\mathcal{Z}_1\big|_{\psi=-\pi/2},\\
S_{(2)}&=\Xi\,\mathcal{Z}_2\big|_{\psi=-\pi/2},\\
T_{(0)}&=\Xi^2\left(u^t_{(0)}\right)^2,\\
T_{(1)}&=2\Xi^2 u^t_{(0)}u^t_{(1)},\\
T_{(2)}&=\Xi^2\left[2u^t_{(0)}u^t_{(2)}+\left(u^t_{(1)}\right)^2\right].
\end{align}

\subsection{Plasma coefficients}

For the refractive index,
\begin{equation}
n^2(r)=\mathcal{N}_0+\delta \mathcal{N}_1+\delta^2 \mathcal{N}_2+\mathcal{O}(\delta^3),
\end{equation}
with
\begin{align}
\mathcal{N}_0&=1-\alpha_0\frac{k}{r^h},\\
\mathcal{N}_1&=-\alpha_1\frac{k}{r^h},\\
\mathcal{N}_2&=-\alpha_2\frac{k}{r^h},
\end{align}
the square-root expansion gives
\begin{align}
n_{(0)}&=\sqrt{\mathcal{N}_0},\\
n_{(1)}&=\frac{\mathcal{N}_1}{2\sqrt{\mathcal{N}_0}},\\
n_{(2)}&=\frac{\mathcal{N}_2}{2\sqrt{\mathcal{N}_0}}-\frac{\mathcal{N}_1^2}{8\mathcal{N}_0^{3/2}}.
\end{align}
Hence
\begin{align}
\hat b_{(0)}&=n_{(0)}b_{(0)},\\
\hat b_{(1)}&=n_{(0)}b_{(1)}+n_{(1)}b_{(0)},\\
\hat b_{(2)}&=n_{(0)}b_{(2)}+n_{(1)}b_{(1)}+n_{(2)}b_{(0)},
\end{align}
and
\begin{align}
\hat{\mathcal{Z}}_0&=u^t_{(0)}-\hat b_{(0)}u^\varphi_{(0)},\\
\hat{\mathcal{Z}}_1&=u^t_{(1)}-\hat b_{(0)}u^\varphi_{(1)}-\hat b_{(1)}u^\varphi_{(0)},\\
\hat{\mathcal{Z}}_2&=u^t_{(2)}-\hat b_{(0)}u^\varphi_{(2)}-\hat b_{(1)}u^\varphi_{(1)}-\hat b_{(2)}u^\varphi_{(0)}.
\end{align}

\bibliography{ref}

\end{document}